\documentclass[prb,twocolumn,aps,showpacs,amsmath,amssymb,superscriptaddress,floatfix]{revtex4-1} 

\usepackage{color}
\usepackage{bm}
\usepackage{graphicx}
\usepackage{braket}

\usepackage[pdftex]{hyperref}
\hypersetup{colorlinks = true, urlcolor = blue, linkcolor = blue, citecolor = blue}

\begin{document}
\title{Fractal x-ray edge problem at the critical point of the Aubry-Andre model}
\author{Ang-Kun Wu}
\affiliation{Department of Physics and Astronomy, Center for Materials Theory, Rutgers University, Piscataway, NJ 08854 USA}
\author{Sarang Gopalakrishnan}
\affiliation{Department of Engineering Science and Physics, CUNY College of Staten Island, Staten Island, NY 10314 USA and Initiative for the Theoretical Sciences, CUNY Graduate Center, New York, NY 10016 USA}
\author{J. H. Pixley}
\affiliation{Department of Physics and Astronomy, Center for Materials Theory, Rutgers University, Piscataway, NJ 08854 USA}
\date{\today}

\begin{abstract}
We study the Anderson orthogonality catastrophe, and the corresponding X-ray edge problem, in systems that are at a localization transition driven by a deterministic quasiperiodic potential. Specifically, we address how the ground state of the Aubry-Andre model, at its critical point, responds to an instantaneous local quench. At this critical point, both the single-particle wavefunctions and the density of states are fractal. We find, numerically, that the overlap between post-quench and pre-quench wavefunctions, as well as the ``core-hole'' Green function, evolve in a complex, non-monotonic way with system size and time respectively. We interpret our results in terms of the fractal density of states at this critical point. In a given sample, as the post-quench time increases, the system resolves increasingly finely spaced minibands, leading to a series of alternating temporal regimes in which the response is flat or algebraically decaying. In addition, the fractal critical wavefunctions give rise to a quench response that varies strongly from site to site across the sample, which produces broad distributions of many-body observables.
Upon averaging this broad distribution over samples, we recover coarse-grained power laws and dynamical exponents characterizing the X-ray edge singularity. We discuss how these features can be probed in ultra-cold atomic gases using radio-frequency spectroscopy and Ramsey interference.
\end{abstract}

\maketitle

\section{Introduction}
\label{sec:into}

Significant advances in our understanding of strongly correlated phenomena have resulted from studying problems that retain inherently non-perturbative many-body effects despite being theoretically tractable. One famous example is the so-called X-ray edge problem, that describes the interaction of a single immobile hole with an electron gas that is introduced instantaneously at some time $t$, Ref.~\onlinecite{Gogolin-2004}. The solution of this problem~\cite{Mahan-1967,Nozieres-1969,Mahan-1988}, which is inherently linked to Anderson's ``orthogonality catastrophe''~\cite{Anderson-1967}, was a key step in understanding the Kondo effect, which in turn underlies our intuitions about a variety of strong-correlation effects, e.g. through the dynamical mean field theory~\cite{RevModPhys.68.13}.

Many standard theoretical approaches to correlated systems begin by modeling electrons in a regular lattice and linearizing the electronic dispersion about the Fermi surface. In one dimension, the Fermi surface consists of two points, so the natural electronic degrees of freedom are essentially left- and right-moving plane waves with an approximately linear dispersion. This is the starting point, e.g., for the powerful bosonization method~\cite{Giamarchi-book}. However, many physically relevant systems are far enough from regular crystalline lattices that this plane-wave assumption is inapplicable. In the limit of strong disorder, a different (and incompatible) toolbox based on the real-space renormalization group becomes controlled~\cite{fisher_singularities}; the most challenging regime is when the single particle wavefunctions are ``critical'' or multifractal, i.e. due to the energy (or coupling constant) being tuned to an Anderson localization quantum phase transition~\cite{Evers-2008}. In disordered systems, the density of states (DOS) evolves smoothly across the Anderson transition; however, in quasicrystals, the DOS also becomes fractal~\cite{kohmoto82}, with parametrically flat bands in which correlation effects are presumably strongly enhanced~\cite{pwhg, magicangle} .

The nature of correlation effects in quasicrystals~\cite{Chalker-2015} is of direct experimental relevance, e.g., to the heavy fermion quasicrystal Au$_{51}$Al$_{34}$Yb$_{15}$, which appears to host a quantum-critical ground state without fine-tuning~\cite{Deguchi-2012}. While, single Kondo impurities have been investigated for a Penrose tiling~\cite{Andrade-2015}, the fundamental nature of the Kondo effect in these systems remains poorly understood. A ``mean field'' picture~\cite{Vlad-1992,Chakravarty-2000,Kettemann-2012} would reduce the problem to a Kondo temperature that is set by the local density of states of the impurity site evaluated at the Fermi energy~\cite{Andrade-2015}. This theory compares well to numerical renormalization group calculations for disordered systems~\cite{Zhuravlev-2007}; 
however, quasicrystals can produce an energy spectrum that is not continuous but is instead fractal~\cite{Ma-1989,Luck-1989}, so it is inappropriate in general to linearize the band structure about the Fermi energy~\cite{Vidal-2001}. In this work we take a different perspective and treat the fractal energy spectrum exactly.
With this in mind, a natural theoretical starting point is to break down the Kondo effect into the ``Coulomb gas'' framework~\cite{Anderson-1970}, which represents the partition function as an  infinite series of spin flips that are coupled (in imaginary time) through the conduction band.  A single spin-flip process interacting with the Fermi sea is precisely captured by the solution of the X-ray edge problem, which itself needs to be reformulated to capture the fractal spectrum and critical wavefunctions. 

While the effects of disorder on the orthogonality catastrophe and X-ray edge singularity has been considered previously~\cite{Chen-1992,Aleiner-1998,Gefen-2002}, recent theoretical work has considered Anderson and many-body localized~\cite{Nandkishore-2015} phases and focused on its statistical nature. As a result, it was
found that there is a non-zero probability that the wavefunction following a local quench has an exponentially vanishing overlap with the initial state~\cite{Khemani-2015,Deng-2015,Deng-2017}.
In contrast, much less is currently known about the nature of the X-ray edge problem and the orthogonality catastrophe when the single particle wavefunctions are critical~\cite{vasseur_moore}. With recent developments in ultracold atomic gases, the X-ray edge spectra can be directly measured using radio-frequency spectroscopy and Ramsey interference~\cite{Knap-2012}. Moreover, ultracold atom setups have demonstrated the ability to emulate the Aubry-Andre model~\cite{Azbel-1979,Aubry-1980} by generating one-dimensional quasiperiodic potentials~\cite{Roati-2008}, thus the X-ray edge spectra composed of multifractal eigenstates can be directly probed experimentally.

In this manuscript, we consider the X-ray edge problem, and the corresponding orthogonality catastrophe, when the spectrum is fractal and the single-particle eigenstates that make up the many-body wavefunction are critical. Here, the single particle states are taken from the critical point of the one-dimensional Aubry-Andre model. 
We find that, on average, the overlap between pre- and post-quench ground state wavefunctions vanishes 
with increasing system size, while acquiring a great deal of structure due to the fractal spectrum. In addition, the distribution of overlaps becomes maximally broad at the critical point. 
We also study the temporal decay of the core-hole Green function, which captures the dynamics of the X-ray edge singularity. 
We find that the fractal gap structure gives rise to a behavior that alternates between an insulating and metallic response in both the wavefunction overlap and the core-hole Green function. 
We therefore define ``coarse grained'' power-law exponents to   
estimate an ``average'' dynamical exponent at the Aubry-Andre critical point. 
The response at late times is very sensitive to the filling: the quasiperiodic potential has hard band gaps at all scales, and our results are dramatically different depending on whether the Fermi energy lies in one of the large band gaps that are resolvable at early times. When the Fermi energy is away from any such band gap, we find a dynamical exponent $z\approx 2$. For a Fermi energy that is near a band gap we find dramatically different behavior: when the Fermi level is in a band gap, there is no orthogonality catastrophe; meanwhile, when the Fermi energy is very near a band edge, the wavefunction overlap is anomalously strongly suppressed because the impurity can create mid-gap localized states, which act effectively as if they were bound states in the standard orthogonality catastrophe~\cite{combescot1971infrared}.

Broad distributions are thus a central feature of this unusual orthogonality catastrophe. To understand how these distributions arise, it is helpful to think of the incommensurate Aubry-Andre potential as a limiting case of a series of periodic rational approximants, with ratio $p/q$ (for instance the ratio  $F_{k-1}/F_k$ where $F_k$ is the $k$th Fibonacci number). The band structure of an approximant consists of $q$ bands, each of bandwidth $\sim 1/q^2$ at the critical point~\cite{kohmoto82}. We now imagine increasing $q$ while keeping the filling (\emph{not} the chemical potential) fixed. The chemical potential then moves through a series of increasingly small band gaps and minibands, so the wavefunction overlap fluctuates between 0 and 1. (The value of the overlap is also highly sensitive to where in the $q$-site unit cell the impurity is located.) One can picture the temporal dynamics analogously: the system resolves a band-gap at scale $q$ on a timescale $\sim q^2$; thus, one can model the core-hole Green function at time $t$ as being captured by the orthogonality catastrophe for an approximant with $q(t) \sim \sqrt{t}$, at a chemical potential that is known only to the same resolution. As $t$ increases, both the band structure and the Fermi level are resolved to increasing precision, and as the Fermi level moves relative to the band structure, the behavior of the core-hole Green function switches between that of an insulator and a metal. This is clearly borne out of our numerical calculations presented in this manuscript, which demonstrates this alternating behavior between  plateaus (representative of an insulator) and a power-law decay (indicative of a metal). In the following, we develop an understanding of this structure by connecting it to the fractal spectrum.

The remainder of the paper is organized as follows: in Sec.~\ref{sec:model} we discuss the model we consider and the methods we use to numerically solve the problem. In Sec.~\ref{sec:results}, we discuss our results on the wavefunction overlap as well as the core-hole Green function and in Sec.~\ref{sec:conclusions} we conclude and discuss the implications of our results.

\section{Model and Approach}
\label{sec:model}

\begin{figure*}
\centering
\includegraphics[width=0.98\textwidth]{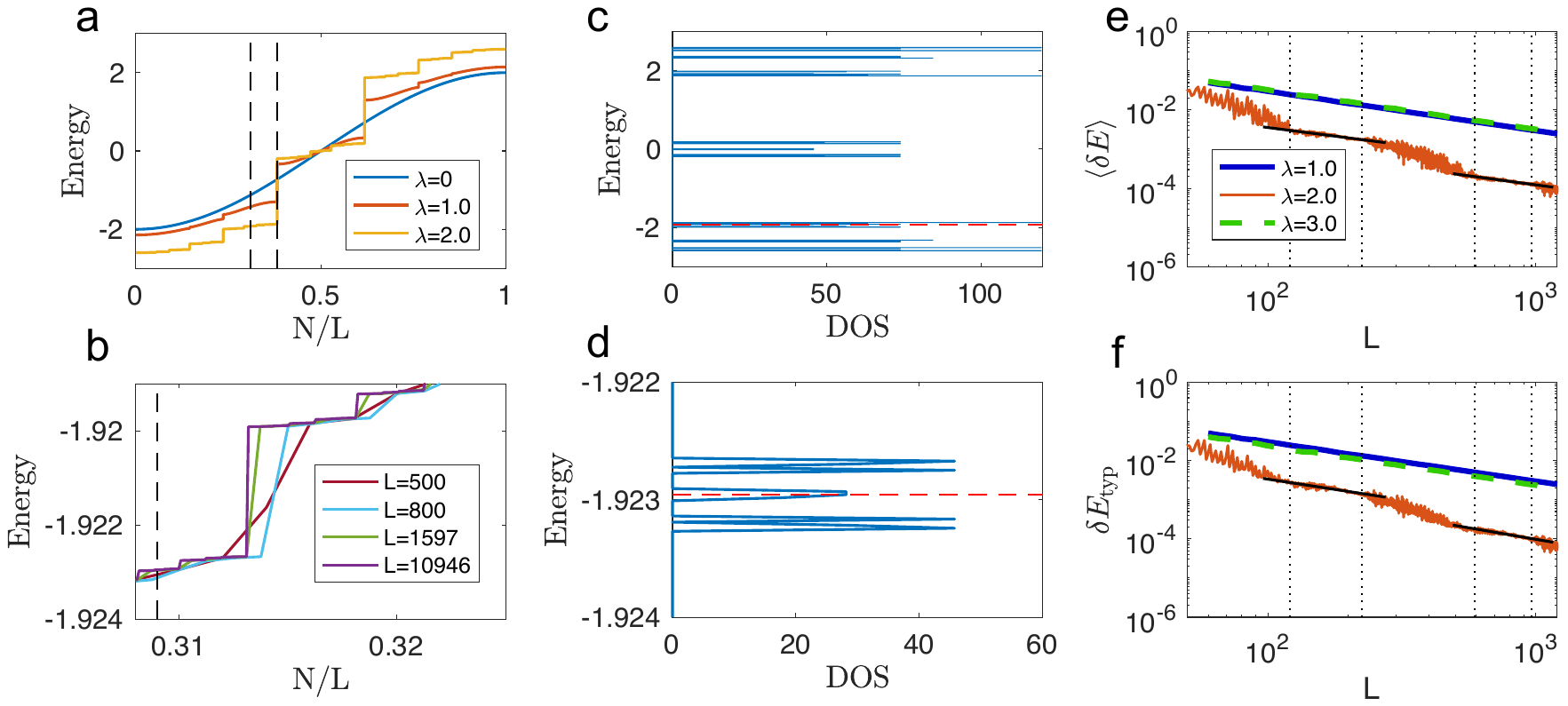}
\caption{\label{fig:filling} Properties of the energy spectrum of the Aubry-Andre model. (a) The energy spectrum of the AA model in different phases (set hopping strength $J=1$),  the black vertical dashed lines mark representative fillings  $n=0.309$ and $0.382$ for a system size $L=10946$. At filling $n=0.309$ and this scale, the Fermi energy sits in a metallic band of the spectrum while filling $n=0.3820=\lim_{k\rightarrow \infty}F_{k-2}/F_k$ is right at a large band gap.  Small steps in each plateau are gaps in the spectrum. (b) The evolution of energy spectrum as the system size $L$ increases. The filling $n=0.309$, that is marked by the black vertical dashed line, develops several minibands as $L$ increases to $10946$. (c) The density of states (DOS) as a function of energy $E$ at the critical point $\lambda_c = 2$ demonstrates the existence of various bands and how our choice of filling appears (the red dashed line).
(d) The DOS for energies very close to the Fermi energy corresponding to the filling of $n=0.309$, which displays the fractal gap structure. This fine resolution shows our choice of filling (the red dashed line) is in a metallic band at this scale. However, due to the fractal spectrum of the problem, more minibands separated by gaps will appear if we zoom in to a much finer energy resolution with larger $L$.  (e) Mean energy difference between the Fermi energy and the first excited state $\langle \delta E \rangle$, for a filling $n=0.309$, as a function of system size $L$, and for coupling constants $\lambda$ in each phase and at the AA critical point ($\lambda_c = 2)$. In each phase $\langle \delta E \rangle$ has a simple power-law scaling, $\delta E\sim L^{-z},z=1$ as shown by the solid blue and dashed green curves. In contrast at the critical point, the fractal spectrum is resolved as a function of $L$: the average $\delta E$ develops structure that crosses gaps (``noisy'' parts) and bands (smooth parts marked by black vertical dotted lines). The critical power laws in band parts are  $z=0.88\pm 0.07$ in the small $L$ regime and $z=0.93 \pm 0.06 $ in the large $L$ regime.  (f) Typical mean $\delta E_{\mathrm{typ}}\equiv \exp\langle \log\delta E \rangle$  for $n=0.309$ as a function of system size $L$. The critical power laws in band parts are $z=1.05\pm 0.10$ for small $L$ part and $z=1.17 \pm 0.05 $ for large $L$ part, respectively.
}
\end{figure*}

We will focus on how the single particle eigenstates of the Aubry-Andre model~\cite{Hofstadter-1976,Azbel-1979,Aubry-1980} are affected by a local quench.
The model Hamiltonian can be written as
\begin{equation}
H_{AA}=-\sum_i J(c_i^\dagger c_{i+1}+\mathrm{H.c.})+\lambda  \cos(2\pi Q i+\phi) c_i^\dagger c_i,
\label{eqn:ham}
\end{equation}
where $J$ is the hopping strength (in the following we take $J=1$ as the unit of energy), $\lambda$ represents the strength of the quasiperiodic potential, $c_i^\dagger$ and $c_i$ are creation and annihilation operators of a spinless fermion at site $i$.  Here, $Q$ is an irrational wave number, e.g., $Q=2/(\sqrt{5}+1)$, and $\phi$ is a randomly chosen phase between 0 and $2\pi$ that is the same for all sites. 
We have considered open boundary conditions, periodic boundary conditions, and twisted boundary conditions. We present results for open boundary conditions, as we find this provides the widest range of acceptable system sizes, while reducing finite size effects. For periodic and twisted boundary conditions we take $Q$ from a rational approximant given by the ratio of Fibonacci numbers, $Q=F_{k-1}/F_k$ and we take the system size to be $L=F_k$.

It is known that all the eigenstates of $H_{AA}$ are localized when $\lambda >2$ and extended when $\lambda <2$  (Ref.~\onlinecite{Aubry-1980}). 
The critical eigenstates ($\lambda = \lambda_c =2$)  are multifractal~\cite{hiramoto1992electronic}. Thus, we can vary $\lambda$ to compare the results for ballistic plane waves ($\lambda < \lambda_c)$ to the case with multifractal wavefunctions at the critical point.
The absence of a mobility edge is beneficial for our purposes, since it allows us to form a many-body wavefunction only out of critical eigenstates.

We take the initial Hamiltonian to be given by Eq.~\eqref{eqn:ham}, $H_I=H_{AA}$ with $N$ particles for time $t<0$. To construct the initial wavefunction, we fill up $N$ single particle states. We will use the fact that the  Hamiltonian can be written as 
$H_I = \sum_{k=1}^N h_k$, where $h_k $ is the single particle Hamiltonian of the $k$th particle. We denote eigenstates of $h$ as $\chi_i(r) = \langle r| \chi_i \rangle$ and energies as $E_i^0$.
At time $t=0$, we quench the system by  introducing a potential scattering term  at one arbitrary site (due to the random phase ($\phi$) it does not matter where we put the quench site), and for convenience we put it at the center of the lattice, i.e. at position $i=L/2$ for a system size $L$. The final Hamiltonian for $t>0$ is given by 
\begin{equation}
H_F=H_{AA}+V_0c_{L/2}^\dagger c_{L/2},
\end{equation}
and we take $V_0=5$ as a representative quench. We will also discuss the value of $V_0=10$ to test the generality of our results, which we present in the Appendix.  The final Hamiltonian can also be written as $H_F = \sum_{k=1}^N \bar h_k$, where 
$\bar h_k $ is the final single-particle Hamiltonian of the $k$th particle. The single-particle eigenstates of  $\bar h$ are denoted as $\psi_j(r) = \langle r | \psi_j \rangle $ with energy $E_j$.
Note that we will not discuss the case for $V_0 <0$, which introduces an additional bound state in the spectrum~\cite{combescot1971infrared}. While this is an interesting effect it will obscure the features that are solely due to a many-body state of multifractal wavefunctions and therefore is not considered here.

\subsection{Choice of filling}
The quasiperiodic potential introduces numerous gaps  in the energy spectrum, and at the critical point this gap structure becomes fractal with a rich mathematical description~\cite{Luck-1989,hiramoto1992electronic}. For example, fixing periodic boundary conditions with $Q = F_{k-1}/F_k$ and a system size $L=F_k$, the spectrum splits into $N_b$ bands and the number of bands scales like $N_b \sim [(\sqrt{5}+1)/2]^k$ (Ref.~\onlinecite{hiramoto1992electronic}). So when we increase $L$, these $N_b$ bands split up even further. If we fix the filling based on data at size $L$, this will eventually land in a ``gap'' at some larger system size $L'$. Therefore, the nature of choosing the filling in this problem is both subtle and important.

Firstly, we only work at fixed filling -- if we instead fix the chemical potential, the system will always lie in a gap at sufficiently large system size due to the fractal spectrum. In other words, for each fixed chemical potential there is a distinct cross over length scale where the singular continuous spectrum goes from ``looking'' gapless to gapped. By working with a fixed filling, on the other hand, we always fill up states to some Fermi energy, the distinction now being whether this Fermi energy lies next to a gap at this $L$ or falls within a miniband.

Away from the critical point, as the potential strength $\lambda$ is varied, some of these gaps can become large while others remain quite small, see Fig.~\ref{fig:filling}. To mimic the X-ray edge problem in a metal, and to study the generic features of the model, our focus is on constructing a gas of particles with multifractal wavefunctions and thus our focus is on $\lambda=\lambda_c$. For this to still resemble a metallic setting, we must ensure that this  corresponds to filling up the single particle states so that the Fermi energy does not lie near any large gap that appears near $\lambda_c$ for the accessible system sizes considered here. Thus, based on Fig.~\ref{fig:filling} we choose a filling $n=N/L=0.309$ (with a Fermi energy $E_F\approx-1.923$). We will also briefly discuss  behavior of the specific case of a Fermi energy lying near a large band gap with $n=0.382$.

Despite choosing $n=0.309$ based on the above criterion, we can still see the effects of fractal gap structure, even though this filling has the advantage that we never reach a large fractal gap up to system sizes of $L=10000$. To see this clearly, we examine the energy difference of the first excited state above the Fermi energy, which is defined as $\delta E = E_{i+1} - E_{i}$ where $E_{i}=E_F$ is the Fermi energy for a given filling $n$ and we average over 10000 samples. If we are in a regime where the spectrum looks continuous, then $ \langle  \delta E \rangle $ will vanish in the large $L$ limit (where $\langle \dots \rangle$ denote an average over the random phases in the quasiperiodic potential). Whereas, if we fill up to the edge of a gap then $ \langle \delta E \rangle$ will be $L$-independent and saturate to a non-zero constant (which does not occur for $n=0.309$ for the system sizes we can reach here). As we can see in Fig.~\ref{fig:filling} in either localized or delocalized phase away from the critical point, there is a clear power-law scaling with $L$, that goes like $\langle \delta E \rangle \sim1/L$. On the other hand, at the critical point the fractal spectrum gives rise to a much richer structure, with a decay in $L$ that is no longer a clear power law. 
As shown in Fig.~\ref{fig:filling}, for $\langle \delta E \rangle$ with $n=0.309$, we find different decay regions with different oscillation amplitudes.  We attribute regimes of large oscillation in $\delta E$ to probing the fractal gaps, whereas the regimes where the oscillations are substantially reduced with a decay that follows 
$\langle \delta E \rangle \sim1/L$ closely, are attributed to the system being well described by being in a band, and hence a ballistic metal like response.
Meanwhile, we find that the typical energy difference $\delta E_{\mathrm{typ}}\equiv \exp\langle \log\delta E \rangle$ decreases with $L$, with a power law slightly larger than the average value. However, this response is very sensitive to our choice of filling and this structure changes if we deviate very slightly away from $n=0.309$ . In order to verify this, as discussed in Sec.~\ref{sec:sens}, instead of trying to increase $L$ even further we consider nearby fillings that reveal a rich fractal gap structure that is strongly dependent on the filling.

This analysis  demonstrates that for various ranges of system sizes, a given filling will go from having a response that looks like it is ``in'' a band to looking like it is filled up to a ``gap''. Since such gaps will always appear at larger and larger system sizes, this process is expected to continue indefinitely in the thermodynamic limit.

\subsection{Evaluating wavefunction overlaps and the core-hole Green function}
We now discuss the general framework we use to numerically compute the wavefunction overlap and the core-hole Green function. While it is well known for this problem that one can write the  many-body wavefunction overlap and core-hole Green function as determinants over single particle eigenstates, for clarity we briefly present this method here following Ref.~\onlinecite{combescot1971infrared}. Since we are focusing on the AA model in Eq.~\eqref{eqn:ham}, we are able to use exact diagonalization on the single particle Hamiltonians to reach sufficiently large system sizes.

We now discuss computing the overlap between the pre- and post-quench many-body wavefunctions $S \equiv \langle \Psi_I | \Psi_F \rangle$.
The many-body fermonic states are defined by the Slater determinant of $N$ single particle eigenstates:
\begin{equation}
\begin{split}
\langle r_1, \dots, r_N \ket{\Psi_I}&=\frac{1}{\sqrt{N!}} \mathrm{det}|\chi_k(r_i)|,\\
\langle r_1, \dots, r_N \ket{\Psi_F}&=\frac{1}{\sqrt{N!}} \mathrm{det}|\psi_{k'}(r_j)|.
\end{split}
\label{eqn:wf}
\end{equation}
As shown in Ref.~\onlinecite{Anderson-1967} the wavefunction overlap $S$ can be written as
\begin{equation}
S = \mathrm{det}|A_{ij}|,
\end{equation}
where $A_{ij}$ is a matrix of overlap integrals of filled electronic states:
\begin{equation}
A_{ij} \equiv \sum_{r}  \psi_i (r) \chi^*_{j}(r) , \quad E_i, E^0_j < E_F. 
\end{equation}
We now come to computing the core-hole Green function $\mathcal{G}(t)$ following Ref.~\onlinecite{combescot1971infrared}.
This is defined as
\begin{equation}
\begin{split}
\mathcal{G}(t)\equiv -\bra{\Psi_I}e^{i\hat{H}_{F} t}e^{-i\hat{H}_{I} t}\ket{\Psi_I}=-e^{-iE_0 t}\bra{\Psi_I}e^{i\hat{H}_{F} t}\ket{\Psi_I},
\end{split}
\label{eqn:G}
\end{equation}
where the initial and final states $\ket{\Psi_I}$ and $\ket{\Psi_F}$ are
given by Eq.~\eqref{eqn:wf}, and $E_0$ is the ground state energy $E_0=\sum_{i \leq N} E^0_i$. In the following we focus on $|\mathcal{G}(t)|$ and therefore, do not need to worry about the phase $\exp(-iE_0t)$.
Expanding the wavefunctions into Slater determinants, we have
\begin{equation}
\begin{split}
\bra{\Psi_0}e^{i\hat{H}_{F} t}\ket{\Psi_0} = \mathrm{det}|\Lambda_{ij}|,\\
\Lambda_{ij}=\bra{\chi_{i}}e^{i\bar{h} t}\ket{\chi_{j}},
\end{split}
\label{eqn:G2}
\end{equation}
$\Lambda_{ij}$ is a one electron matrix element, and $\bar h$ is the single particle Hamiltonian after the quench.

We have now transformed the many-body calculation into one that only involves single-particle matrix elements \cite{combescot1971infrared}, which we can evaluate numerically on system sizes up to $L=10946$ and thus capture the multifractal nature of the wavefunctions non-perturbatively. 

\subsection{Expectations from the plane-wave case}
\label{sec3d}

We conclude this section by briefly reviewing established results for the plane-wave case~\cite{Mahan-2013}, to which we will be comparing our fractal results. The canonical orthogonality catastrophe (or X-ray edge) problem involves a Fermi energy far from the band edge and the core hole produces a scattering phase shift $\delta$. In this case, we have that $\mathcal{G}(t) \sim t^{-2(\delta/\pi)^2}$; the overlap decays with a similar exponent $S(L) \sim L^{-(\delta/\pi)^2}$. The relation between these two exponents comes from the dynamical critical exponent $z = 1$ for a Fermi liquid, in which space and time are related by the Fermi velocity $v_F$. In a finite-size system, the temporal decay of the Green function stops when $t \sim L$; past this time, the Green function saturates to $S(L)^2$ (corresponding to the diagonal-ensemble prediction, which applies after complete dephasing). 

When the Fermi energy lies in a large band gap, the overlap is size-independent, and the Green function saturates on a timescale of order unity. For small band-gaps, one expects on physical grounds that this timescale should be proportional to the band gap, and the overlap should correspondingly decrease; however, we are not aware of previous systematic studies of this dependence. In one dimension, if one adds attractive interactions and the Fermi level is near the bottom of a band (or repulsive interactions when the Fermi level is near the top of the band), the impurity potential creates a bound state, which leads to a strong suppression of the overlap (since, at the one-particle level, the overlap between a plane wave and a bound state is $\sim 1/\sqrt{L}$). 

\section{Results}
\label{sec:results}

We now come to our findings on the wavefunction overlap and the core-hole Green function. As our analysis of the fractal gap's influence on the Fermi energy implies, the wavefunction overlap will go from a metal-like response to that of insulator depending on the system size. Since the metallic regime is well understood and previously outlined in Sec.~\ref{sec3d}, we begin by discussing the effects of choosing a filling close to a band gap.

\subsection{Filling near a large band gap, $n \approx 0.382$}

\begin{figure}[b!]
\begin{center}
\includegraphics[width = 0.4\textwidth]{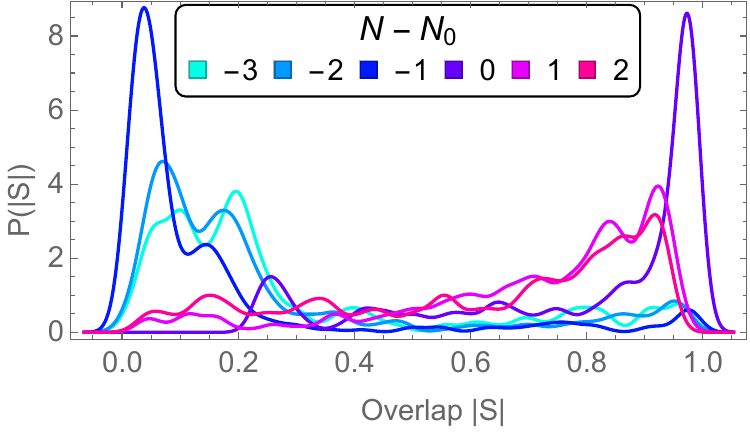}
\includegraphics[width = 0.4\textwidth]{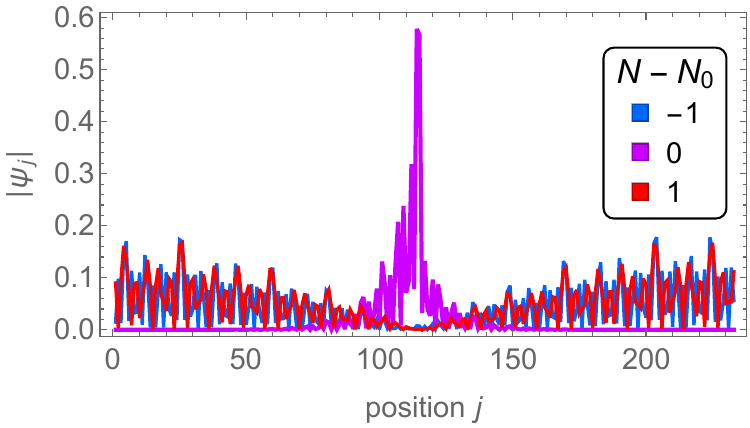}
\caption{Dependence of the wavefunction and its pre and post quench overlap $S$ on the number of particles $N$ near a large band gap that occurs for the filling $n \approx f_0$ (where $f_0 \approx 0.382$ is defined in the text); $N$ states are occupied where $N \approx N_0 = f_0 L$. Upper panel: Distribution of the wavefunction overlap $P[S]$ as $N$ is tuned through $N_0$, at $\lambda = 2$, which is sampled over the phase $\phi$ of the quasiperiodic potential in Eq.~\eqref{eqn:ham}. When the band is fully filled $N = N_0$ the overlap is anomalously large; when $N = N_0 - 1$ the overlap is anomalously small (but this does not occur when $N = N_0 + 1$). 
Lower panel: Absolute value of the single particle wavefunction $|\psi_j|$ as a function of the position $j$ for three different states near the band gap at $N=N_0$.
When $N = N_0 - 1$ the impurity creates an unoccupied (i.e. hole) mid-gap state. When $N = N_0 - 1$ the
pre- and post-quench wavefunctions 
both have one hole in the miniband, but the hole is localized in the unquenched case due to the bound state and delocalized otherwise. As a result, for $N=N_0-1$, the overlap scales as $1/\sqrt{L}$ and is therefore greatly suppressed. To make the contrast between localized and delocalized states clear, the lower panel shows data for $\lambda = 1.75$, slightly in the delocalized phase.}
\label{neargap}
\end{center}
\end{figure}

\begin{figure}[t!]
\centering
\includegraphics[width=0.48\textwidth]{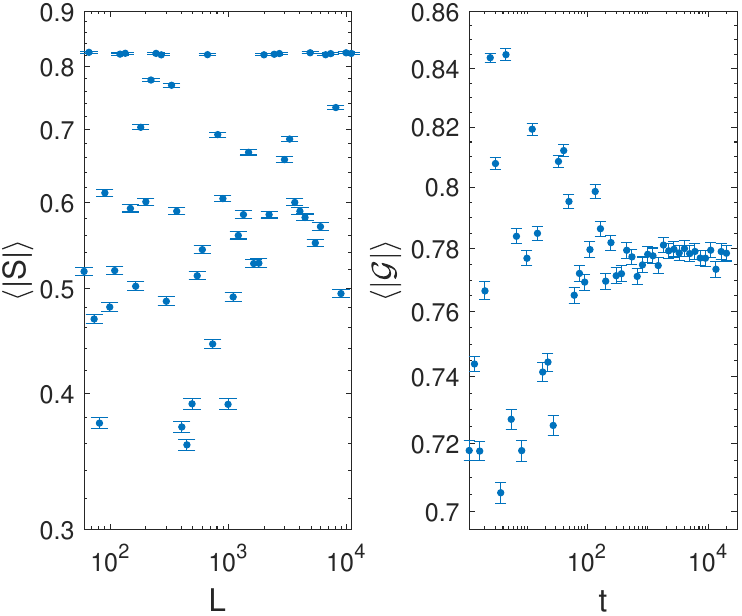}
\caption{\label{fig:Gt38} Results for a filling at the edge of a large band gap, $n=0.382$.  Left panel:  The mean overlap does not exhibit clear decay here, instead it is oscillating and has a clear upper bound due to the large band gap. Right panel: The core-hole Green function saturates to a non-zero value in the long-time limit for a system size $L=1597$. }
\end{figure}

As shown in Fig.~\ref{fig:filling}, a large band gap arises in the quasiperiodic band structure for filling $n = f_0$ where $f_0 = \lim_{k \rightarrow \infty} F_k / F_{k+2} \approx 0.382$. This band gap exists when $\lambda \agt 1$. As one tunes the filling through this band gap, the overlap first becomes anomalously small (when the band is nearly filled) and then anomalously large (when the band is fully filled); see Fig.~\ref{neargap}. A large overlap for a filled band is expected (see Sec.~\ref{sec3d}), consistent with what we find in Fig.~\ref{fig:Gt38}. The anomalous suppression comes about because the impurity potential introduces a mid-gap bound state. When the band is nearly filled, one can think of the problem as containing a low density of holes. One of these holes occupies the impurity-bound state when there is an impurity, but a delocalized state otherwise, as Fig.~\ref{neargap} displays. The many-body overlap is dominated in this regime by the single-particle overlap between the bound and delocalized states, and thus scales as $1/\sqrt{N}$, which decreases much faster with $N$ than the usual orthogonality exponent. This leads to anomalous suppression when the filling is just below the band edge, but not when it is just above, as shown in Fig.~\ref{neargap}. This asymmetry is because the impurity potential is repulsive: if it were attractive, the bound state would be particle-like rather than hole-like and the anomalous suppression would be for fillings slightly above a band-edge. In either case, the rest of our discussion would go through as before. In Fig.~\ref{fig:Gt38} we show representative results for the average overlap and the core-hole Green function for the filling $n=0.382$. We find that filling up to a band gap produces an overlap that is large and oscillating about a mean value, while the core-hole Green function saturates to a non-zero value at long times.

We will argue in what follows that the essential physics of the orthogonality catastrophe at the Aubry-Andre critical point arises from this rapid fluctuation of overlaps with the position of the Fermi level relative to a band gap, together with the proliferation of band gaps on all scales.

\begin{figure}[t!]
\centering
\includegraphics[width=0.45\textwidth]
{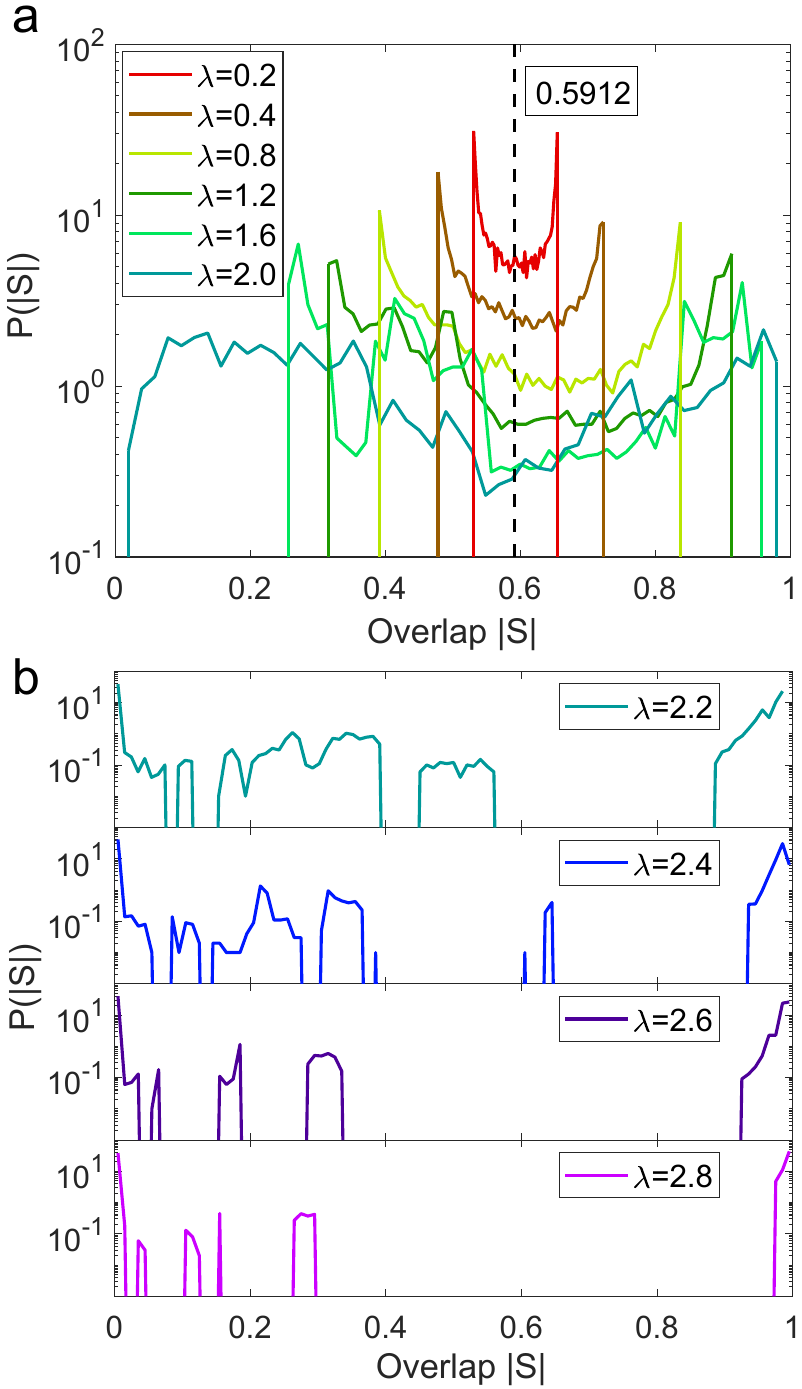}
\caption{\label{fig2:DistOLFS} Distribution of overlap integral for a fixed system size $L=1096$ as a function of the quasiperiodic potential strength $\lambda$. (a) Overlap in extended phase and at the critical point. The black dashed line marking $\lambda=0$. We find for weak $\lambda$ the distribution develops two peaks towards small and large overlap centered about the value at $\lambda=0$. This distribution widens for increasing $\lambda$ until it becomes  broad at the critical point, stretching across all possible values of the overlap. (b) Overlap in the Anderson localized phase with bin size $0.01$. The distribution is no longer broad and develops two peaks centered about 0 and 1, which reflects the existence of the statistical orthogonality catastrophe.  }
\end{figure}

\subsection{Wavefunction overlap $S$}

We begin by discussing the distribution of wavefunction overlaps as a function of $\lambda$ starting in the plane-wave limit $(\lambda=0)$. For a finite system size the overlap at $\lambda=0$ will be non-zero (taking values between zero and one), therefore we find it convenient to mark it as a vertical dashed  line. As shown in Fig.~\ref{fig2:DistOLFS}, as  $\lambda$ increases from zero, the distribution of wavefunction overlaps $P[S]$ develops two peaks towards large and small overlap, while centering around the $\lambda=0$ value. As $\lambda$ approaches the critical value ($\lambda=2$), $P[S]$ continues to broaden, and the two peaks separately approach zero and one. Importantly, at the critical point $P[S]$ has become maximally broad with weight at all values stretching from zero to one. Upon entering the localized phase $(\lambda>2)$, the statistical orthogonality catastrophe leads to peaks in $P[S]$ about zero and one with the vanishing weights at intermediate overlap values. 

\begin{figure}[b!]
\centering
\includegraphics[width=0.45\textwidth]{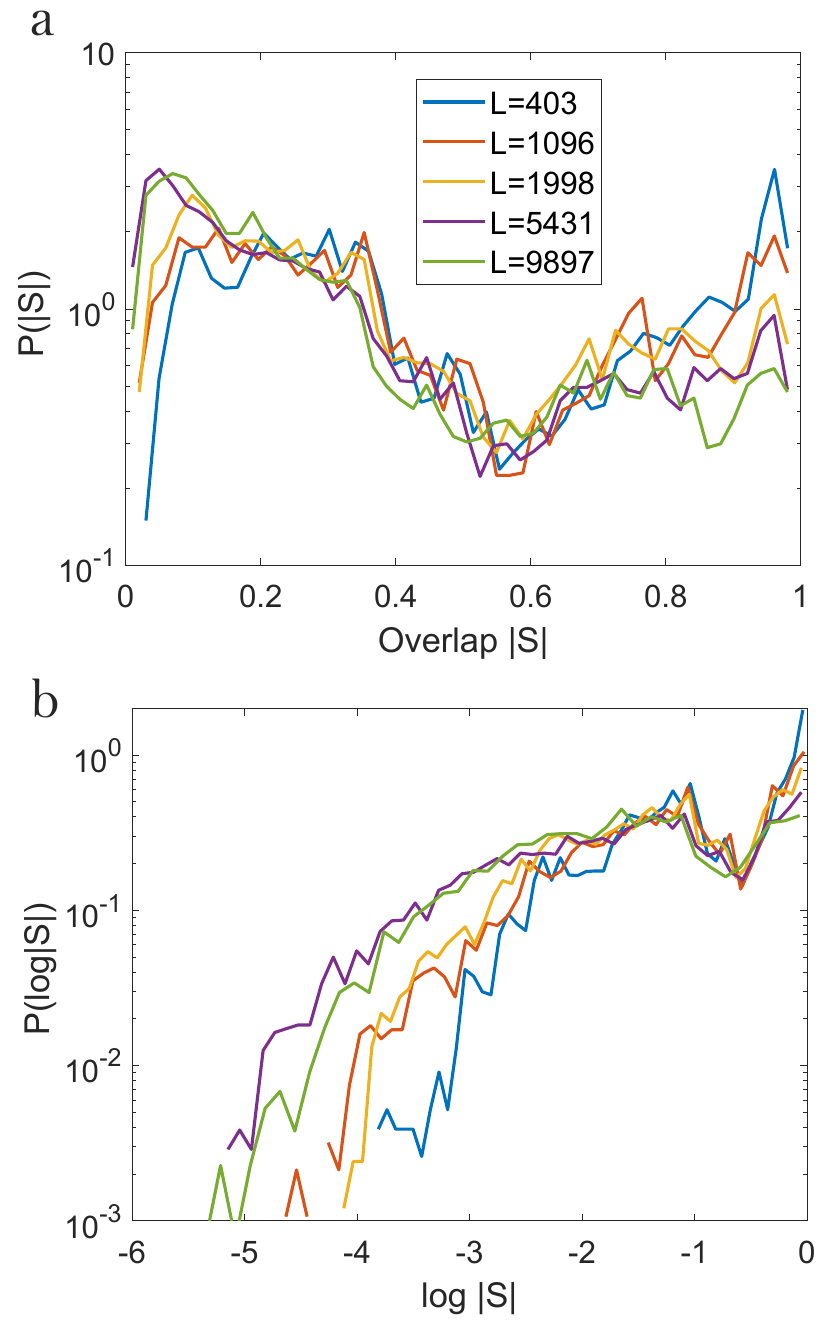}
\caption{\label{fig:DistOLFS} Distribution of the wavefunction overlap  $|S|$ and $\log|S|$ for five different system sizes, focusing on the critical point $\lambda=2$. (a) The distribution of $|S|$ spread over $0$ to $1$. The general shape is similar across different sizes. (b) The distribution of $\log|S|$ displaying a clear tail towards vanishing overlap that develops for  large $L$  and a decrease around $|S|=1$. }
\end{figure}

To probe the distribution  
at the critical point and the broad nature of $P[S]$ we turn to the finite size dependence as shown in Fig.~\ref{fig:DistOLFS}. 
As $S \rightarrow 0$, we find $P[S]$ develops more weight towards zero and a significant tail towards vanishing overlap that becomes remarkably broad at large $L$, stretching across 5 decades.  On the other hand, for $S\rightarrow 1$, $P[S]$ also decreases with increasing $L$. These results suggest that the \emph{average} wavefunction overlap will vanish with increasing $L$. This contrasts with the localized phase that has a non-vanishing average overlap, despite the typical overlap vanishing in the large $L$ limit~\cite{Deng-2015}.

\begin{figure}[h!]
\centering
\includegraphics[width=0.48\textwidth]{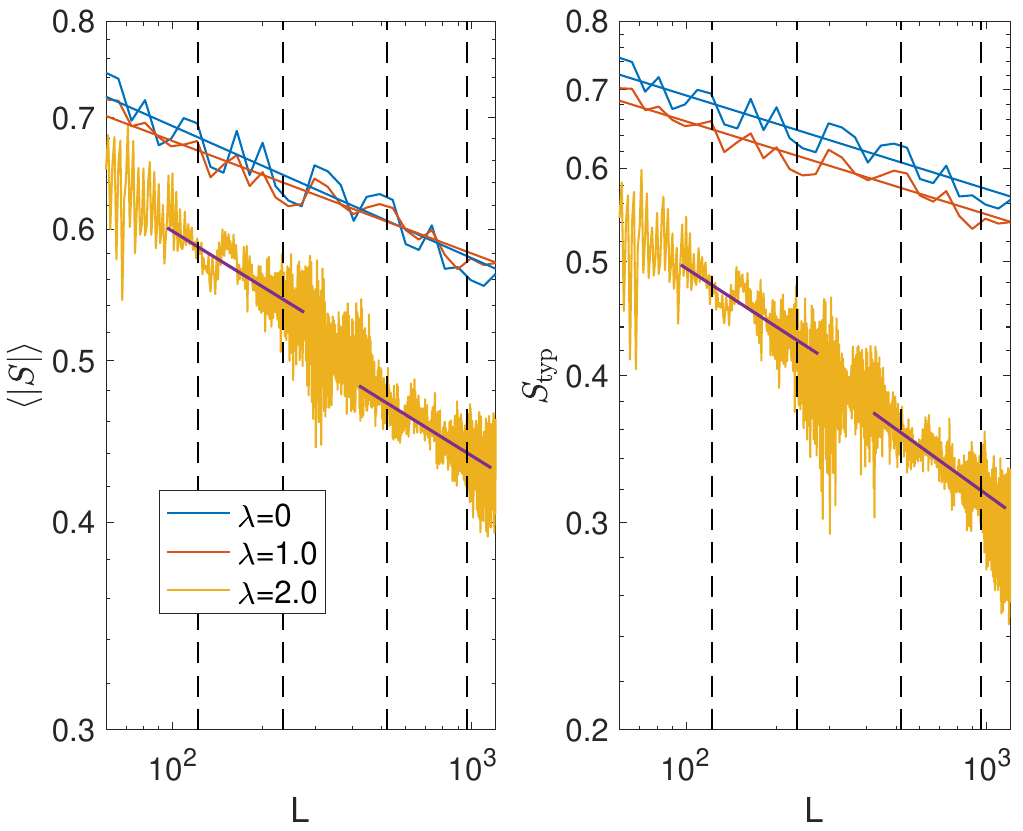}
\caption{\label{fig:L0} System size dependence of the average and typical  overlap $|S|$  for various values of $\lambda$  averaged over $10^4$ realizations with open boundary conditions and filling $n=0.309$.  The lines are the linear fit corresponding to each power-law decay. Left panel: The power-law decays 
extracted from the fits are $\gamma=0.08\pm 0.006$ for $\lambda=0$, $\gamma=0.07 \pm 0.004$ for $\lambda = 1$, 
$\gamma=0.11\pm 0.03 $ for $\lambda=2$ in the first power-law regime, and $\gamma=0.11\pm 0.03 $ in the second. Right panel: For the power-law decay in the typical overlap (note that $\lambda=0$ is equivalent to the left panel as there is no quasiperiodic potential) we find $\gamma_{\mathrm{typ}}=0.08 \pm 0.005$ for $\lambda=1$, $\gamma_{\mathrm{typ}}=0.17 \pm 0.04 $ for $\lambda=2$, in the first power-law regime, and  $\gamma_{\mathrm{typ}}=0.18 \pm 0.02 $ in the second. In the localized phase ($\lambda>2$), the typical overlap $S_{\mathrm{typ}}$ decays exponentially~\cite{Deng-2015}. }
\end{figure}

To see how these results modify the conventional orthogonality catastrophe we consider the system size ($L$) dependence of the average and typical overlap, defined as $S_{\mathrm{avg}} = \langle | S| \rangle$ and $S_{\mathrm{typ}}=\exp{\langle  \log |S| \rangle}$ respectively, where $\langle \dots \rangle$ denotes an average over the random phases in the Aubry-Andre model. 
As displayed in Fig.~\ref{fig:L0}, in the metallic phase ($\lambda < 2$) we find a clear power law decay in the average and typical overlap. Whereas at the critical point  ($\lambda=2$)  we find that the average wavefunction overlap decays but develops an oscillatory behavior as well. As we increase $L$, we go from regimes in $\langle \delta E \rangle$ (see Fig.~\ref{fig:filling}) that are highly oscillatory (due to the fractal spectrum) to a regime that looks like it is in a band and hence metallic. Thus,  to correctly extract the power-law decay we have to restrict the system sizes to those that have a metallic like response in $\langle \delta E \rangle$, which provides us with the vertical dashed lines in Fig.~\ref{fig:L0}. 
In addition, we also find that the typical overlap, which probes the weight of the  tail towards vanishing overlap, vanishes with a power law different from the average, namely
\begin{equation}
S_{\mathrm{avg}} \sim L^{-\gamma}, \,\,\,\,\, S_{\mathrm{typ}} \sim L^{-\gamma_{\mathrm{typ}}},
\end{equation}
where the power-law exponents are not expected to be universal but depend on the value of the local potential $V_0$ and filling $n$. For $V_0=5$ and $n=0.309$, we find in the first power-law regime $\gamma \approx 0.11$ and $\gamma_{\mathrm{typ}} \approx 0.17$ and in the second power-law regime we find $\gamma \approx 0.11 $ and $\gamma_{\mathrm{typ}} \approx 0.18 $, which implies that the exponents governing each metallic regime are the same within our numerical accuracy. This typical value is somewhat larger than what we find for the same potential strength away from the critical point, and is close to the unitary limit (as one would expect, since the vanishing bandwidth at criticality implies that any scattering potential is in effect a strong one). 

Eventually for larger $L$, in a regime where $\langle \delta E \rangle$ will have a plateau in $L$, the power-law behavior of $S$ changes. For these larger sizes we expect the results will instead resemble that of a gapped regime. Here, the overlap oscillates but is no-longer clearly vanishing with $L$ (akin to the results in Fig.~\ref{fig:Gt38}). Thus, we expect that for $L$ increasing arbitrarily, the overlap will continuously go back and forth between these regimes, i.e., a regime where the overlap decreases algebraically with $L$ and another in which it does not decay but oscillates strongly about a ``mean'' value. This is consistent with results at nearby fillings (as opposed to trying to increase $L$ any further) that have a very different plateau-decay structure, as discussed in Sec.~\ref{sec:sens}.

\begin{figure}
\centering
\includegraphics[width=0.48\textwidth]{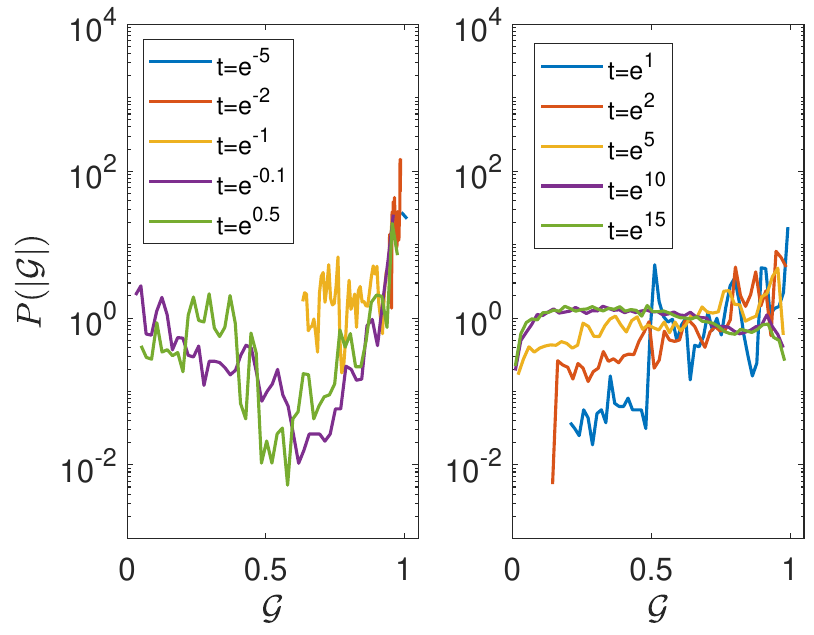}
\caption{\label{fig:Dist31} Distribution of the time-dependent core-hole Green function at the critical point $\lambda=2$ and at the filling $n=0.309$. We show the time-dependent overlap function in system size $L=1597$. Initially centering at $1$, the distribution of the Green function evolves into broad structure and saturates in the long-time limit due to finite size effect. The saturated distribution resembles the distribution of ground state overlap.  }
\end{figure}

\subsection{Core-hole Green function $\mathcal{G}(t)$}

We now turn to the time dependence of the wavefunction overlap, which is captured by computing the core-hole Green function $\mathcal{G}(t)$. In the following, we focus on $| \mathcal{G}(t) | =| \langle \Psi_0(t) |  \Psi_0 \rangle|$,  which is defined in Eq.~\eqref{eqn:G}. Similar to $S$, we  focus on the distribution of the core-hole Green function $P[\mathcal{G}(t)]$ and study how it evolves dynamically as a function of time.

We find that $P[\mathcal{G}]$ develops a broad distribution in the long-time limit.  This is shown clearly in Fig.~\ref{fig:Dist31}, where the weight in the distribution at short times is concentrated near $\mathcal{G}=1$ and vanishes towards small $\mathcal{G}$. For increasing $t$, the small $\mathcal{G}$ tail fills in
and $P[\mathcal{G}]$ becomes almost uniformly distributed 
as  $t\rightarrow \infty$.  At long times we find that the distribution becomes $t$-independent, the time scale at which this occurs is set by the finite system size.

\begin{figure}[h!]
\centering
\includegraphics[width=0.48\textwidth]{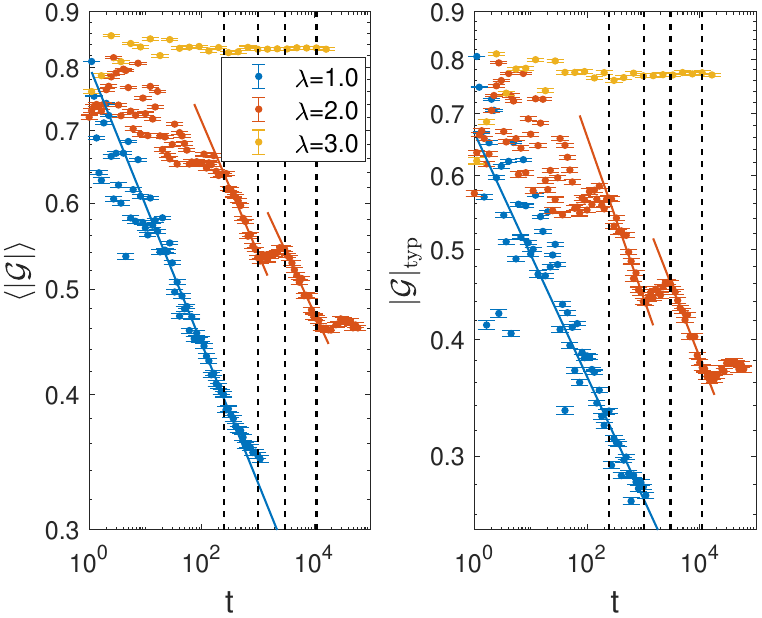}
\caption{\label{fig:Gt31} Average and typical $|\mathcal{G}|$ at filling $n=0.309$ for a system size $L=1597$ and the representative behavior in each phase.  The mean $\langle |\mathcal{G}(t)| \rangle$ and typical $\mathcal{G}_{\mathrm{typ}}(t)$ and their error bars are calculated over $10^4$ samples with open boundary conditions. The blue, red and yellow curves are at $\lambda=1, 2,$ and 3, respectively. The dashed black lines mark different decay and plateau regions and the straight lines are fits to a power law form for  $\lambda=1$ and 2. 
In the delocalized phase we find a clear power-law decay, whereas in the Anderson insulating phase the core-hole Green function saturates to a time independent non-zero value, reminiscent of a filling up to a band edge (see Fig.~\ref{fig:Gt38}).
Left panel: Power-law decays in $\langle |\mathcal{G}(t)| \rangle$ extracted from the fits over the appropriate time regimes  are $\beta=0.13\pm 0.01$ for $\lambda=1$, $\beta=0.12 \pm 0.01$ for $\lambda=2$ in the first regime of decay, and $\beta=0.11 \pm 0.01$ for $\lambda=2$ in the second decay regime. Right panel: Power-law decays in $\mathcal{G}(t)_{\mathrm{typ}}$  extracted from the fits are $ \beta_{\mathrm{typ}} = 0.13 \pm 0.02$ for $\lambda=1$, $\beta_{\mathrm{typ}} =0.17 \pm 0.02$ for $\lambda=2$ in the first decay regime and $\beta_{\mathrm{typ}}=0.16 \pm 0.02 $ for $\lambda= 2$ in the second regime of decay. }
\end{figure}

To see how this behavior manifests itself in the conventional X-ray edge response, we turn to the average and typical core-hole Green function, defined as $\mathcal{G}_{\mathrm{avg}}(t)=\langle | \mathcal{G}(t)| \rangle$ and $\mathcal{G}_{\mathrm{typ}}(t)=\exp\left(\langle \log| \mathcal{G}(t)| \rangle\right)$, respectively. In general, as discussed in the introduction, we find that the Green function evolves through a series of alternating power-law decays and plateaus, which is clearly displayed in the long-time regime in Fig.~\ref{fig:Gt31}. Similar to our analysis of the overlap, we restrict fitting the data to regimes that have a clear power-law decay in $t$. For this filling ($n=0.309$) we find two clear power-law regimes separated by a small plateau, which in each power law regime yields 
\begin{equation}
\mathcal{G}_{\mathrm{avg}}(t) \sim t^{-\beta}, \,\,\,\,\, \mathcal{G}_{\mathrm{typ}}(t) \sim t^{-\beta_{\mathrm{typ}}}.
\end{equation}
Similar to $S$, the power-law exponents are not expected to be universal and will depend on the value of the local potential $V_0$ and filling $n$. For $V_0=5$ and $n=0.309$
we find $\beta \approx 0.12$ and $\beta_{\mathrm{typ}} \approx 0.17$ in the first power-law regime as well as $\beta \approx 0.11$ and $\beta_{\mathrm{typ}} \approx 0.16$ in the second regime. Within our numerical accuracy, we find the power laws in the two distinct power-law regimes coincide.   In each power-law regime, the  average power-law decay is markedly distinct from the delocalized phase. This is expected as transport is not ballistic at the critical point, so the dynamical critical exponent $z > 1$, which we turn to in the following section. 

In the Appendix we compare our results for the exponents $\beta$ and $\beta_{\mathrm{typ}}$ for two different quench potentials. $V_0=5$ and $V_0=10$.
Interestingly, at the critical point ($\lambda=2$) we find that the power-law exponents in the  core-hole Green function are unaffected by increasing the strength of the quench potential, which implies the system is close to the unitary limit. Whereas, in the metallic phase $(\lambda < 2$) we find the exponents strongly depend on the value of $V_0$ and are therefore still far away from the unitary scattering regime. We attribute this effect to the vanishing bandwidth from the fractal gap structure that   enhances the strength of correlations and hence the effective strength of $V_0$.

\subsection{Coarse grained dynamical exponent}
At conventional critical points, the dynamic critical exponent $z$ describes the power-law scaling between energy and length via $E \sim L^{-z}$. 
For the orthogonality catastrophe, we can extract $z$ from our data on the scaling of $S(L)$ and $\mathcal{G}(t)$, where $z = 2\gamma/\beta$ and $z_{\mathrm{typ}} = 2\gamma_{\mathrm{typ}}/\beta_{\mathrm{typ}}$, see Sec.~\ref{sec3d}.
However, as we have clearly seen throughout this manuscript, the fractal gap structure complicates any conventional power laws relating length and energy (similar to what was found in Ref.~\onlinecite{Schneider-2018}). This is demonstrated clearly in our results for $\delta E$ in Figs.~\ref{fig:filling} and \ref{fig:widefig}, not satisfying a simple power law across the full range of $L$. In contrast, in the delocalized phase, taking $\lambda=1$ as a representative case we find $z = 1.1 \pm 0.1$ and $z_{\mathrm{typ}} = 1.2 \pm 0.2$, in good agreement with the expectation of a ballistic metal.

Despite the fractal gap structure at the critical point, we can provide an estimate of a coarse grained or an ``averaged'' dynamical exponent from fitting the data across a regime in $L$ and a corresponding regime in $t$ at which the system encounters no gaps. We have done this in Figs.~\ref{fig:L0} and \ref{fig:Gt31}. For the two power-law regimes that we have accessed to, we find $z=1.8 \pm 0.3$ and $z_{\mathrm{typ}}=2.0 \pm 0.4$ in the first power-law regime as well as $z=2.0 \pm 0.3$ and $z_{\mathrm{typ}}=2.3 \pm 0.4$ in the second power-law regime. Thus, we find that the dynamical response in the metallic scaling regimes (i.e. when the filling is nicely within a band) is diffusive with $z\approx 2$ and consistent across each respective power-law regime. This is consistent with the understanding~\cite{kohmoto82} that the energy-length scaling at the Aubry-Andre critical point has $z = 2$.

\subsection{Sensitivity to filling fraction}
\label{sec:sens}

\begin{figure}[t]
\begin{center}
\includegraphics[width=0.48\textwidth]{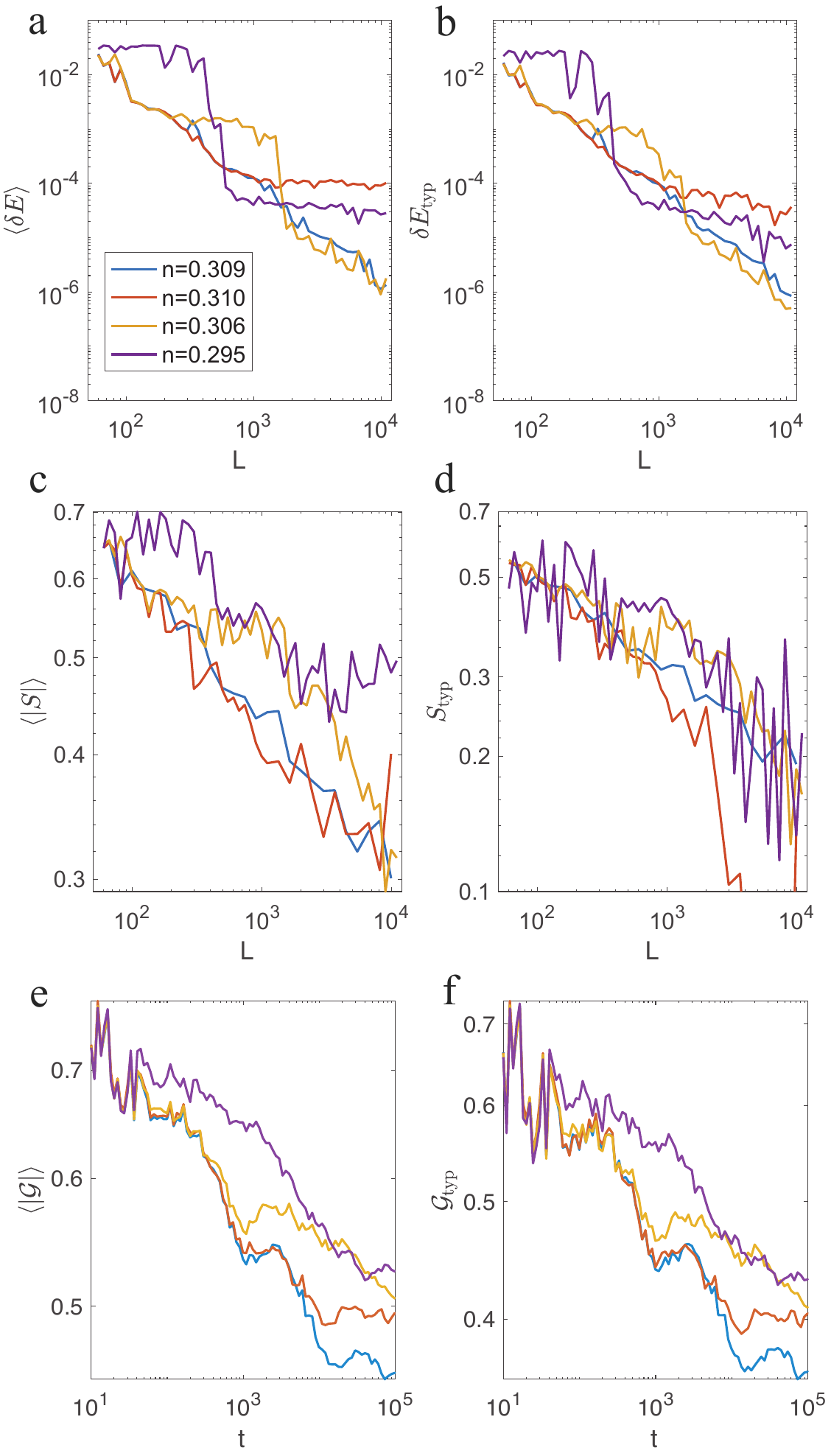}
\caption{Comparison between four representative fillings around $n=0.309$. The mean and typical energy difference between the Fermi energy and the first excited state are shown in (a) and (b). The corresponding results for the mean and typical overlap $S$ are shown in (c) and (d) as well as the mean and typical Green function $\mathcal{G}$ (for $L=1597$) are shown in (e) and (f).  }
\label{fig:widefig}
\end{center}
\end{figure}

In the previous sections, we focused on a long algebraic segment of the decay in order to extract exponents. However, as Fig.~\ref{fig:widefig} shows, a striking feature of the fractal orthogonality catastrophe is that these algebraic segments are interrupted by plateaus in $\langle \delta E \rangle$ signaling the appearance of a fractal gap. In these plateaus, the Fermi level is at the edge of a band (to within the resolution given by system size and/or time). A plateau ends when the resolution is increased further, to the point that the system now no longer has a filled band. This structure of repeated decays and plateaus is clear in the data (Fig.~\ref{fig:widefig}): note that plateaus in the gap (i.e., level spacing) at the Fermi energy, the overlap, and the core-hole Green function track each other for all the fillings considered. The correlation between these measures points to the central role that the fractal density of states plays in the quasiperiodic orthogonality catastrophe. This cements our picture of the fractal X-ray edge problem probing a sequence of metallic and insulating regimes of response that alternates indefinitely in the thermodynamic limit.

\section{Discussion and Conclusions}
\label{sec:conclusions}

In this work we addressed the behavior of the Anderson orthogonality catastrophe and the X-ray edge singularity at the critical point of the Aubry-Andre model. This critical point is unusual in having not only fractal wavefunctions (which are generically present at localization transitions) but also a fractal density of states. This latter feature, which is central to understanding our results, does not exist in random systems but is quite generic in quasiperiodic ones: it exists, essentially by construction, in many common models of quasicrystals that are defined by dilation rules. It is also possibly relevant to the band structure of physical quasicrystals in higher dimensions, which are of increasing experimental relevance. In addition, following the exciting experimental discoveries of twisted bilayer graphene~\cite{Cao-2018}, understanding the interplay of incommensurate effects on electronic structure and strong correlations is of increasing theoretical importance.

As one approaches the critical point of the Aubry-Andre model from the delocalized side, more and more band gaps open up; correspondingly the bandwidth of each band decreases. There are three types of behaviors. (1) When the Fermi energy is in the middle of a band, one has a fairly conventional orthogonality catastrophe, with a phase shift that increases toward the critical point (since the kinetic energy decreases). (2) When the Fermi energy is in a band gap, the wavefunction overlap does not scale with system size, corresponding to no orthogonality catastrophe. (3) When the Fermi energy is very close to a band edge, the orthogonality catastrophe is enhanced, and the overlap decreases as $L^{-1/2}$, because of the influence of mid-gap bound states. Away from criticality, scenario (1) is generic. However, at the critical point, there are band gaps at all scales, so depending on the impurity position and the precise filling, essentially any value of the overlap between 0 and 1 can arise. We found numerically that the overlap and core-hole Green function (i.e., Loschmidt echo) have regimes of algebraic decay, consistent with a dynamical critical exponent $z \approx 2 $, as well as flat regimes, in $L$ and $t$, such that the band structure is gapped at the available spatio-temporal resolution. 
 The precise alternation of gaps and decays is sensitive to the filling (Fig.~\ref{fig:widefig}). This dependence originates in number-theoretic considerations that are outside the scope of this work. The qualitative pattern, however, can be seen by considering a series of approximants with increasingly large denominator $q$. At a fixed filling $n$, the number of electrons in the system is $\lfloor n q\rfloor$ (i.e. the closest lower integer to $nq$). For fillings where $n q$ is close to a Fibonacci number (or the sum of a few Fibonacci numbers) the system is gapped; whereas when $n q$ can only be expressed as a sum of many ($\sim \log q$) Fibonacci numbers, the response decays with $L$ algebraically. Similar results hold upon replacing system size with time, and account for the response of the core-hole Green function. 
 
Note that this structure relies on working at fixed filling rather than fixed chemical potential: in the latter case, the critical point is generically gapped, and there is no orthogonality catastrophe. The assumption of fixed filling is sensible for metallic quasicrystals, since the filling in these will be fixed by the chemical composition of the compounds. On the other hand, fixed filling will generically be difficult to achieve in ultracold atomic gases (except in box traps), since the harmonic trapping potential gives a spatially varying chemical potential, which will likely render the system locally gapped. However, square-bottomed optical traps are increasingly common in ultracold atomic experiments, and these allow one to work at fixed particle number~\cite{zoran2013}.

The orthogonality catastrophe is perhaps the most basic manifestation of the physics of strong correlations; as we have seen, it is profoundly modified by the fractal band structure that is common in quasicrystals. A natural extension of this work would be to study the Kondo effect and its generalizations, as well as to derive effective models for quantum magnetism, in fractal band structures~\cite{Chalker-2015}. A further question, raised by the large susceptibility of these flat-band systems and their extreme sensitivity to slight changes in filling, is how robust they are against phase separation in the presence of even weak interactions.

\begin{acknowledgments}
We acknowledge useful discussions with Natan Andrei, Piers Coleman, Sriram Ganeshan, Gabriel Kotliar, Sid Parameswaran, Qimiao Si, and Justin Wilson.
S.G. acknowledges support from PSC-CUNY Grant No. 61656-00 49. S.G. and J.H.P. performed part of this work at the Aspen Center for Physics, which is supported by NSF Grant No. PHY-1607611, and at the Kavli Institute for Theoretical Physics, which is supported by NSF Grant No. PHY-1748958.  
The authors acknowledge the Beowulf cluster at the Department of Physics and Astronomy of Rutgers University and the Office of Advanced Research Computing (OARC) at Rutgers, The State University of New Jersey (http://oarc.rutgers.edu) for providing access to the Amarel cluster and associated research computing resources that have contributed to the results reported here.
\end{acknowledgments}

\appendix*
\section{Dependence on $V_0$}
\begin{figure}[h!]
\centering
\includegraphics[width=0.48\textwidth]{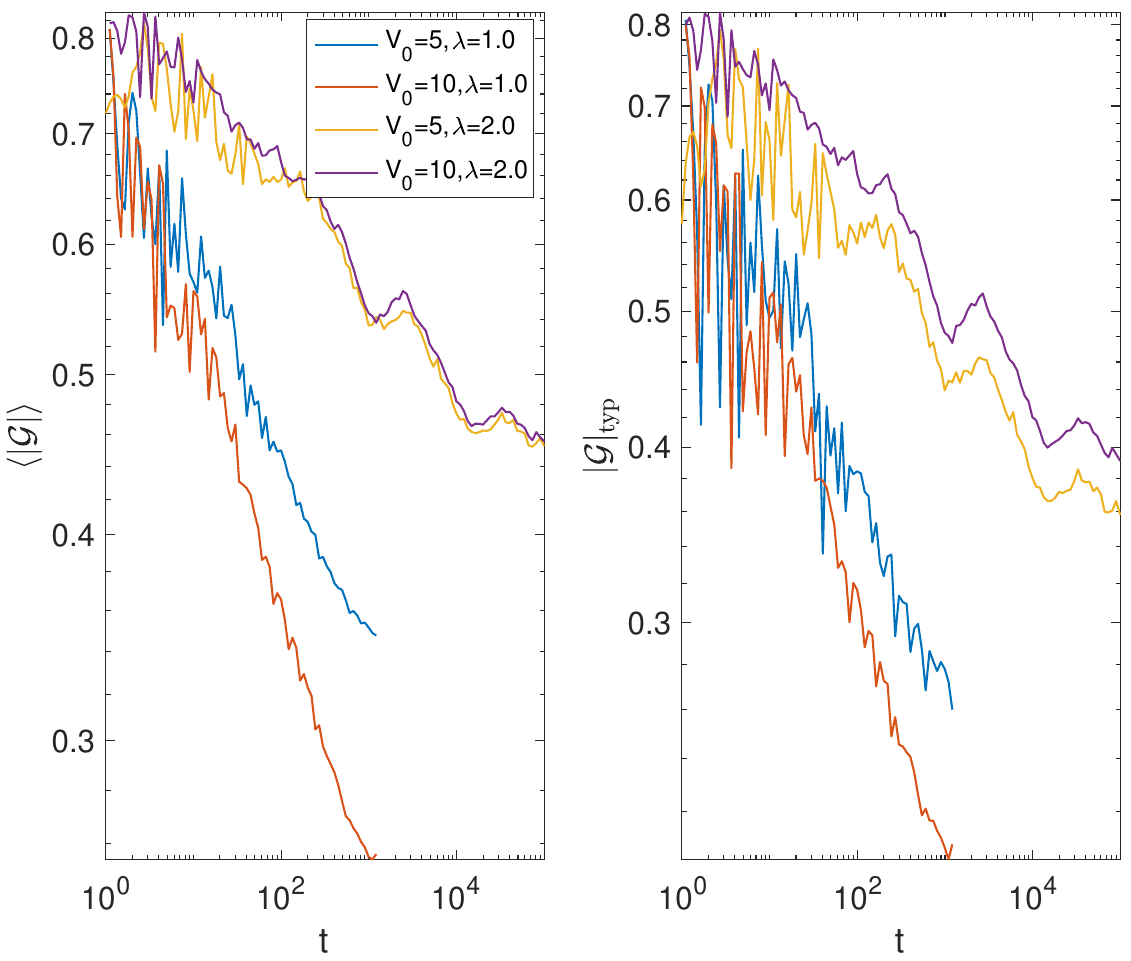}
\caption{\label{fig:Gt31Vs} Average and typical $|\mathcal{G}|$ at filling $n=0.309$ for a system size $L=1597$ comparing two different values of the quench potential $V_0$ in the delocalized phase and at the critical point. Note that for $\lambda=1$ the end of the power-law regime is due to the finite system size. 
 Left panel: The exponents results for $\mathcal{G}_{\mathrm{avg}}(t) $ with $V_0=10$ are:
for $\lambda=1$,  $\beta=0.17 \pm 0.01$; for $\lambda=2$ in the first power-law regime $\beta=0.12 \pm 0.01$ and in the second regime $\beta=0.11 \pm 0.01$. Right panel: The results for typical means $\mathcal{G}_{\mathrm{typ}}(t)$ with $V_0=10$ are: for $\lambda = 1$, $\beta_{\mathrm{typ}}=0.2 \pm 0.01$; for $\lambda=2$ in the first region $\beta_{\mathrm{typ}}=0.17 \pm 0.02$ and in the second region $\beta_{\mathrm{typ}}=0.15 \pm 0.03$.  }
\end{figure}
We briefly discuss the dependence of our results on the choice of the quench potential $V_0$. In the main text we focused on a potential strength of $V_0=5$, we now compare this for the core-hole Green function for $V_0=10$. In the delocalized phase these values of $V_0$ are not large enough to be in the unitary limit, i.e. they do not  reach the largest possible value of the scattering length and thus the exponents are distinct between $V_0 = 5$ and $V_0 = 10$. In contrast, at the critical point, due to the strongly renormalized bandwidth of each miniband, correlation effects are strongly enhanced, and we find the power-law response in $\mathcal{G}(t)$ is essentially identical between $V_0= 5$ and $V_0 = 10$.

\bibliography{QP_AOC}

\begin{thebibliography}{42}%
\makeatletter
\providecommand \@ifxundefined [1]{%
 \@ifx{#1\undefined}
}%
\providecommand \@ifnum [1]{%
 \ifnum #1\expandafter \@firstoftwo
 \else \expandafter \@secondoftwo
 \fi
}%
\providecommand \@ifx [1]{%
 \ifx #1\expandafter \@firstoftwo
 \else \expandafter \@secondoftwo
 \fi
}%
\providecommand \natexlab [1]{#1}%
\providecommand \enquote  [1]{``#1''}%
\providecommand \bibnamefont  [1]{#1}%
\providecommand \bibfnamefont [1]{#1}%
\providecommand \citenamefont [1]{#1}%
\providecommand \href@noop [0]{\@secondoftwo}%
\providecommand \href [0]{\begingroup \@sanitize@url \@href}%
\providecommand \@href[1]{\@@startlink{#1}\@@href}%
\providecommand \@@href[1]{\endgroup#1\@@endlink}%
\providecommand \@sanitize@url [0]{\catcode `\\12\catcode `\$12\catcode
  `\&12\catcode `\#12\catcode `\^12\catcode `\_12\catcode `\%12\relax}%
\providecommand \@@startlink[1]{}%
\providecommand \@@endlink[0]{}%
\providecommand \url  [0]{\begingroup\@sanitize@url \@url }%
\providecommand \@url [1]{\endgroup\@href {#1}{\urlprefix }}%
\providecommand \urlprefix  [0]{URL }%
\providecommand \Eprint [0]{\href }%
\providecommand \doibase [0]{http://dx.doi.org/}%
\providecommand \selectlanguage [0]{\@gobble}%
\providecommand \bibinfo  [0]{\@secondoftwo}%
\providecommand \bibfield  [0]{\@secondoftwo}%
\providecommand \translation [1]{[#1]}%
\providecommand \BibitemOpen [0]{}%
\providecommand \bibitemStop [0]{}%
\providecommand \bibitemNoStop [0]{.\EOS\space}%
\providecommand \EOS [0]{\spacefactor3000\relax}%
\providecommand \BibitemShut  [1]{\csname bibitem#1\endcsname}%
\let\auto@bib@innerbib\@empty
\bibitem [{\citenamefont {Gogolin}\ \emph {et~al.}(2004)\citenamefont
  {Gogolin}, \citenamefont {Nersesyan},\ and\ \citenamefont
  {Tsvelik}}]{Gogolin-2004}%
  \BibitemOpen
  \bibfield  {author} {\bibinfo {author} {\bibfnamefont {A.~O.}\ \bibnamefont
  {Gogolin}}, \bibinfo {author} {\bibfnamefont {A.~A.}\ \bibnamefont
  {Nersesyan}}, \ and\ \bibinfo {author} {\bibfnamefont {A.~M.}\ \bibnamefont
  {Tsvelik}},\ }\href@noop {} {\emph {\bibinfo {title} {Bosonization and
  strongly correlated systems}}}\ (\bibinfo  {publisher} {Cambridge University
  Press},\ \bibinfo {year} {2004})\BibitemShut {NoStop}%
\bibitem [{\citenamefont {Mahan}(1967)}]{Mahan-1967}%
  \BibitemOpen
  \bibfield  {author} {\bibinfo {author} {\bibfnamefont {G.~D.}\ \bibnamefont
  {Mahan}},\ }\href {\doibase 10.1103/PhysRev.163.612} {\bibfield  {journal}
  {\bibinfo  {journal} {Phys. Rev.}\ }\textbf {\bibinfo {volume} {163}},\
  \bibinfo {pages} {612} (\bibinfo {year} {1967})}\BibitemShut {NoStop}%
\bibitem [{\citenamefont {Nozieres}\ and\ \citenamefont
  {De~Dominicis}(1969)}]{Nozieres-1969}%
  \BibitemOpen
  \bibfield  {author} {\bibinfo {author} {\bibfnamefont {P.}~\bibnamefont
  {Nozieres}}\ and\ \bibinfo {author} {\bibfnamefont {C.~T.}\ \bibnamefont
  {De~Dominicis}},\ }\href {\doibase 10.1103/PhysRev.178.1097} {\bibfield
  {journal} {\bibinfo  {journal} {Phys. Rev.}\ }\textbf {\bibinfo {volume}
  {178}},\ \bibinfo {pages} {1097} (\bibinfo {year} {1969})}\BibitemShut
  {NoStop}%
\bibitem [{\citenamefont {Mahan}(1988)}]{Mahan-1988}%
  \BibitemOpen
  \bibfield  {author} {\bibinfo {author} {\bibfnamefont {G.~D.}\ \bibnamefont
  {Mahan}},\ }in\ \href@noop {} {\emph {\bibinfo {booktitle} {Fermi surface
  effects}}}\ (\bibinfo  {publisher} {Springer},\ \bibinfo {year} {1988})\ pp.\
  \bibinfo {pages} {41--61}\BibitemShut {NoStop}%
\bibitem [{\citenamefont {Anderson}(1967)}]{Anderson-1967}%
  \BibitemOpen
  \bibfield  {author} {\bibinfo {author} {\bibfnamefont {P.~W.}\ \bibnamefont
  {Anderson}},\ }\href {\doibase 10.1103/PhysRevLett.18.1049} {\bibfield
  {journal} {\bibinfo  {journal} {Phys. Rev. Lett.}\ }\textbf {\bibinfo
  {volume} {18}},\ \bibinfo {pages} {1049} (\bibinfo {year}
  {1967})}\BibitemShut {NoStop}%
\bibitem [{\citenamefont {Georges}\ \emph {et~al.}(1996)\citenamefont
  {Georges}, \citenamefont {Kotliar}, \citenamefont {Krauth},\ and\
  \citenamefont {Rozenberg}}]{RevModPhys.68.13}%
  \BibitemOpen
  \bibfield  {author} {\bibinfo {author} {\bibfnamefont {A.}~\bibnamefont
  {Georges}}, \bibinfo {author} {\bibfnamefont {G.}~\bibnamefont {Kotliar}},
  \bibinfo {author} {\bibfnamefont {W.}~\bibnamefont {Krauth}}, \ and\ \bibinfo
  {author} {\bibfnamefont {M.~J.}\ \bibnamefont {Rozenberg}},\ }\href {\doibase
  10.1103/RevModPhys.68.13} {\bibfield  {journal} {\bibinfo  {journal} {Rev.
  Mod. Phys.}\ }\textbf {\bibinfo {volume} {68}},\ \bibinfo {pages} {13}
  (\bibinfo {year} {1996})}\BibitemShut {NoStop}%
\bibitem [{\citenamefont {Giamarchi}(2003)}]{Giamarchi-book}%
  \BibitemOpen
  \bibfield  {author} {\bibinfo {author} {\bibfnamefont {T.}~\bibnamefont
  {Giamarchi}},\ }\href@noop {} {\emph {\bibinfo {title} {Quantum physics in
  one dimension}}}\ (\bibinfo  {publisher} {Clarendon Press},\ \bibinfo {year}
  {2003})\BibitemShut {NoStop}%
\bibitem [{\citenamefont {Fisher}(1999)}]{fisher_singularities}%
  \BibitemOpen
  \bibfield  {author} {\bibinfo {author} {\bibfnamefont {D.~S.}\ \bibnamefont
  {Fisher}},\ }\href@noop {} {\bibfield  {journal} {\bibinfo  {journal}
  {Physica A.}\ }\textbf {\bibinfo {volume} {263}},\ \bibinfo {pages} {222}
  (\bibinfo {year} {1999})}\BibitemShut {NoStop}%
\bibitem [{\citenamefont {Evers}\ and\ \citenamefont
  {Mirlin}(2008)}]{Evers-2008}%
  \BibitemOpen
  \bibfield  {author} {\bibinfo {author} {\bibfnamefont {F.}~\bibnamefont
  {Evers}}\ and\ \bibinfo {author} {\bibfnamefont {A.~D.}\ \bibnamefont
  {Mirlin}},\ }\href {\doibase 10.1103/RevModPhys.80.1355} {\bibfield
  {journal} {\bibinfo  {journal} {Rev. Mod. Phys.}\ }\textbf {\bibinfo {volume}
  {80}},\ \bibinfo {pages} {1355} (\bibinfo {year} {2008})}\BibitemShut
  {NoStop}%
\bibitem [{\citenamefont {Kohmoto}(1983)}]{kohmoto82}%
  \BibitemOpen
  \bibfield  {author} {\bibinfo {author} {\bibfnamefont {M.}~\bibnamefont
  {Kohmoto}},\ }\href {\doibase 10.1103/PhysRevLett.51.1198} {\bibfield
  {journal} {\bibinfo  {journal} {Phys. Rev. Lett.}\ }\textbf {\bibinfo
  {volume} {51}},\ \bibinfo {pages} {1198} (\bibinfo {year}
  {1983})}\BibitemShut {NoStop}%
\bibitem [{\citenamefont {Pixley}\ \emph {et~al.}(2018)\citenamefont {Pixley},
  \citenamefont {Wilson}, \citenamefont {Huse},\ and\ \citenamefont
  {Gopalakrishnan}}]{pwhg}%
  \BibitemOpen
  \bibfield  {author} {\bibinfo {author} {\bibfnamefont {J.~H.}\ \bibnamefont
  {Pixley}}, \bibinfo {author} {\bibfnamefont {J.~H.}\ \bibnamefont {Wilson}},
  \bibinfo {author} {\bibfnamefont {D.~A.}\ \bibnamefont {Huse}}, \ and\
  \bibinfo {author} {\bibfnamefont {S.}~\bibnamefont {Gopalakrishnan}},\ }\href
  {\doibase 10.1103/PhysRevLett.120.207604} {\bibfield  {journal} {\bibinfo
  {journal} {Phys. Rev. Lett.}\ }\textbf {\bibinfo {volume} {120}},\ \bibinfo
  {pages} {207604} (\bibinfo {year} {2018})}\BibitemShut {NoStop}%
\bibitem [{\citenamefont {Fu}\ \emph {et~al.}(2018)\citenamefont {Fu},
  \citenamefont {K{\"o}nig}, \citenamefont {Wilson}, \citenamefont {Chou},\
  and\ \citenamefont {Pixley}}]{magicangle}%
  \BibitemOpen
  \bibfield  {author} {\bibinfo {author} {\bibfnamefont {Y.}~\bibnamefont
  {Fu}}, \bibinfo {author} {\bibfnamefont {E.~J.}\ \bibnamefont {K{\"o}nig}},
  \bibinfo {author} {\bibfnamefont {J.~H.}\ \bibnamefont {Wilson}}, \bibinfo
  {author} {\bibfnamefont {Y.-Z.}\ \bibnamefont {Chou}}, \ and\ \bibinfo
  {author} {\bibfnamefont {J.~H.}\ \bibnamefont {Pixley}},\ }\href@noop {}
  {\bibfield  {journal} {\bibinfo  {journal} {arXiv preprint arXiv:1809.04604}\
  } (\bibinfo {year} {2018})}\BibitemShut {NoStop}%
\bibitem [{\citenamefont {Thiem}\ and\ \citenamefont
  {Chalker}(2015)}]{Chalker-2015}%
  \BibitemOpen
  \bibfield  {author} {\bibinfo {author} {\bibfnamefont {S.}~\bibnamefont
  {Thiem}}\ and\ \bibinfo {author} {\bibfnamefont {J.~T.}\ \bibnamefont
  {Chalker}},\ }\href {\doibase 10.1103/PhysRevB.92.224409} {\bibfield
  {journal} {\bibinfo  {journal} {Phys. Rev. B}\ }\textbf {\bibinfo {volume}
  {92}},\ \bibinfo {pages} {224409} (\bibinfo {year} {2015})}\BibitemShut
  {NoStop}%
\bibitem [{\citenamefont {Deguchi}\ \emph {et~al.}(2012)\citenamefont
  {Deguchi}, \citenamefont {Matsukawa}, \citenamefont {Sato}, \citenamefont
  {Hattori}, \citenamefont {Ishida}, \citenamefont {Takakura},\ and\
  \citenamefont {Ishimasa}}]{Deguchi-2012}%
  \BibitemOpen
  \bibfield  {author} {\bibinfo {author} {\bibfnamefont {K.}~\bibnamefont
  {Deguchi}}, \bibinfo {author} {\bibfnamefont {S.}~\bibnamefont {Matsukawa}},
  \bibinfo {author} {\bibfnamefont {N.~K.}\ \bibnamefont {Sato}}, \bibinfo
  {author} {\bibfnamefont {T.}~\bibnamefont {Hattori}}, \bibinfo {author}
  {\bibfnamefont {K.}~\bibnamefont {Ishida}}, \bibinfo {author} {\bibfnamefont
  {H.}~\bibnamefont {Takakura}}, \ and\ \bibinfo {author} {\bibfnamefont
  {T.}~\bibnamefont {Ishimasa}},\ }\href@noop {} {\bibfield  {journal}
  {\bibinfo  {journal} {Nat. Mater.}\ }\textbf {\bibinfo {volume} {11}},\
  \bibinfo {pages} {1013} (\bibinfo {year} {2012})}\BibitemShut {NoStop}%
\bibitem [{\citenamefont {Andrade}\ \emph {et~al.}(2015)\citenamefont
  {Andrade}, \citenamefont {Jagannathan}, \citenamefont {Miranda},
  \citenamefont {Vojta},\ and\ \citenamefont
  {Dobrosavljevi\ifmmode~\acute{c}\else \'{c}\fi{}}}]{Andrade-2015}%
  \BibitemOpen
  \bibfield  {author} {\bibinfo {author} {\bibfnamefont {E.~C.}\ \bibnamefont
  {Andrade}}, \bibinfo {author} {\bibfnamefont {A.}~\bibnamefont
  {Jagannathan}}, \bibinfo {author} {\bibfnamefont {E.}~\bibnamefont
  {Miranda}}, \bibinfo {author} {\bibfnamefont {M.}~\bibnamefont {Vojta}}, \
  and\ \bibinfo {author} {\bibfnamefont {V.}~\bibnamefont
  {Dobrosavljevi\ifmmode~\acute{c}\else \'{c}\fi{}}},\ }\href {\doibase
  10.1103/PhysRevLett.115.036403} {\bibfield  {journal} {\bibinfo  {journal}
  {Phys. Rev. Lett.}\ }\textbf {\bibinfo {volume} {115}},\ \bibinfo {pages}
  {036403} (\bibinfo {year} {2015})}\BibitemShut {NoStop}%
\bibitem [{\citenamefont {Dobrosavljevi\ifmmode~\acute{c}\else \'{c}\fi{}}\
  \emph {et~al.}(1992)\citenamefont {Dobrosavljevi\ifmmode~\acute{c}\else
  \'{c}\fi{}}, \citenamefont {Kirkpatrick},\ and\ \citenamefont
  {Kotliar}}]{Vlad-1992}%
  \BibitemOpen
  \bibfield  {author} {\bibinfo {author} {\bibfnamefont {V.}~\bibnamefont
  {Dobrosavljevi\ifmmode~\acute{c}\else \'{c}\fi{}}}, \bibinfo {author}
  {\bibfnamefont {T.~R.}\ \bibnamefont {Kirkpatrick}}, \ and\ \bibinfo {author}
  {\bibfnamefont {B.~G.}\ \bibnamefont {Kotliar}},\ }\href {\doibase
  10.1103/PhysRevLett.69.1113} {\bibfield  {journal} {\bibinfo  {journal}
  {Phys. Rev. Lett.}\ }\textbf {\bibinfo {volume} {69}},\ \bibinfo {pages}
  {1113} (\bibinfo {year} {1992})}\BibitemShut {NoStop}%
\bibitem [{\citenamefont {Chakravarty}\ and\ \citenamefont
  {Nayak}(2000)}]{Chakravarty-2000}%
  \BibitemOpen
  \bibfield  {author} {\bibinfo {author} {\bibfnamefont {S.}~\bibnamefont
  {Chakravarty}}\ and\ \bibinfo {author} {\bibfnamefont {C.}~\bibnamefont
  {Nayak}},\ }\href@noop {} {\bibfield  {journal} {\bibinfo  {journal} {Int. J.
  Mod. Phys. B}\ }\textbf {\bibinfo {volume} {14}},\ \bibinfo {pages} {1421}
  (\bibinfo {year} {2000})}\BibitemShut {NoStop}%
\bibitem [{\citenamefont {Kettemann}\ \emph {et~al.}(2012)\citenamefont
  {Kettemann}, \citenamefont {Mucciolo}, \citenamefont {Varga},\ and\
  \citenamefont {Slevin}}]{Kettemann-2012}%
  \BibitemOpen
  \bibfield  {author} {\bibinfo {author} {\bibfnamefont {S.}~\bibnamefont
  {Kettemann}}, \bibinfo {author} {\bibfnamefont {E.~R.}\ \bibnamefont
  {Mucciolo}}, \bibinfo {author} {\bibfnamefont {I.}~\bibnamefont {Varga}}, \
  and\ \bibinfo {author} {\bibfnamefont {K.}~\bibnamefont {Slevin}},\ }\href
  {\doibase 10.1103/PhysRevB.85.115112} {\bibfield  {journal} {\bibinfo
  {journal} {Phys. Rev. B}\ }\textbf {\bibinfo {volume} {85}},\ \bibinfo
  {pages} {115112} (\bibinfo {year} {2012})}\BibitemShut {NoStop}%
\bibitem [{\citenamefont {Zhuravlev}\ \emph {et~al.}(2007)\citenamefont
  {Zhuravlev}, \citenamefont {Zharekeshev}, \citenamefont {Gorelov},
  \citenamefont {Lichtenstein}, \citenamefont {Mucciolo},\ and\ \citenamefont
  {Kettemann}}]{Zhuravlev-2007}%
  \BibitemOpen
  \bibfield  {author} {\bibinfo {author} {\bibfnamefont {A.}~\bibnamefont
  {Zhuravlev}}, \bibinfo {author} {\bibfnamefont {I.}~\bibnamefont
  {Zharekeshev}}, \bibinfo {author} {\bibfnamefont {E.}~\bibnamefont
  {Gorelov}}, \bibinfo {author} {\bibfnamefont {A.~I.}\ \bibnamefont
  {Lichtenstein}}, \bibinfo {author} {\bibfnamefont {E.~R.}\ \bibnamefont
  {Mucciolo}}, \ and\ \bibinfo {author} {\bibfnamefont {S.}~\bibnamefont
  {Kettemann}},\ }\href {\doibase 10.1103/PhysRevLett.99.247202} {\bibfield
  {journal} {\bibinfo  {journal} {Phys. Rev. Lett.}\ }\textbf {\bibinfo
  {volume} {99}},\ \bibinfo {pages} {247202} (\bibinfo {year}
  {2007})}\BibitemShut {NoStop}%
\bibitem [{\citenamefont {Ma}\ and\ \citenamefont {Liu}(1989)}]{Ma-1989}%
  \BibitemOpen
  \bibfield  {author} {\bibinfo {author} {\bibfnamefont {P.}~\bibnamefont
  {Ma}}\ and\ \bibinfo {author} {\bibfnamefont {Y.}~\bibnamefont {Liu}},\
  }\href {\doibase 10.1103/PhysRevB.39.10658} {\bibfield  {journal} {\bibinfo
  {journal} {Phys. Rev. B}\ }\textbf {\bibinfo {volume} {39}},\ \bibinfo
  {pages} {10658} (\bibinfo {year} {1989})}\BibitemShut {NoStop}%
\bibitem [{\citenamefont {Luck}(1989)}]{Luck-1989}%
  \BibitemOpen
  \bibfield  {author} {\bibinfo {author} {\bibfnamefont {J.~M.}\ \bibnamefont
  {Luck}},\ }\href {\doibase 10.1103/PhysRevB.39.5834} {\bibfield  {journal}
  {\bibinfo  {journal} {Phys. Rev. B}\ }\textbf {\bibinfo {volume} {39}},\
  \bibinfo {pages} {5834} (\bibinfo {year} {1989})}\BibitemShut {NoStop}%
\bibitem [{\citenamefont {Vidal}\ \emph {et~al.}(2001)\citenamefont {Vidal},
  \citenamefont {Mouhanna},\ and\ \citenamefont {Giamarchi}}]{Vidal-2001}%
  \BibitemOpen
  \bibfield  {author} {\bibinfo {author} {\bibfnamefont {J.}~\bibnamefont
  {Vidal}}, \bibinfo {author} {\bibfnamefont {D.}~\bibnamefont {Mouhanna}}, \
  and\ \bibinfo {author} {\bibfnamefont {T.}~\bibnamefont {Giamarchi}},\ }\href
  {\doibase 10.1103/PhysRevB.65.014201} {\bibfield  {journal} {\bibinfo
  {journal} {Phys. Rev. B}\ }\textbf {\bibinfo {volume} {65}},\ \bibinfo
  {pages} {014201} (\bibinfo {year} {2001})}\BibitemShut {NoStop}%
\bibitem [{\citenamefont {Anderson}\ \emph {et~al.}(1970)\citenamefont
  {Anderson}, \citenamefont {Yuval},\ and\ \citenamefont
  {Hamann}}]{Anderson-1970}%
  \BibitemOpen
  \bibfield  {author} {\bibinfo {author} {\bibfnamefont {P.~W.}\ \bibnamefont
  {Anderson}}, \bibinfo {author} {\bibfnamefont {G.}~\bibnamefont {Yuval}}, \
  and\ \bibinfo {author} {\bibfnamefont {D.~R.}\ \bibnamefont {Hamann}},\
  }\href {\doibase 10.1103/PhysRevB.1.4464} {\bibfield  {journal} {\bibinfo
  {journal} {Phys. Rev. B}\ }\textbf {\bibinfo {volume} {1}},\ \bibinfo {pages}
  {4464} (\bibinfo {year} {1970})}\BibitemShut {NoStop}%
\bibitem [{\citenamefont {Chen}\ and\ \citenamefont {Kroha}(1992)}]{Chen-1992}%
  \BibitemOpen
  \bibfield  {author} {\bibinfo {author} {\bibfnamefont {Y.}~\bibnamefont
  {Chen}}\ and\ \bibinfo {author} {\bibfnamefont {J.}~\bibnamefont {Kroha}},\
  }\href {\doibase 10.1103/PhysRevB.46.1332} {\bibfield  {journal} {\bibinfo
  {journal} {Phys. Rev. B}\ }\textbf {\bibinfo {volume} {46}},\ \bibinfo
  {pages} {1332} (\bibinfo {year} {1992})}\BibitemShut {NoStop}%
\bibitem [{\citenamefont {Aleiner}\ and\ \citenamefont
  {Matveev}(1998)}]{Aleiner-1998}%
  \BibitemOpen
  \bibfield  {author} {\bibinfo {author} {\bibfnamefont {I.~L.}\ \bibnamefont
  {Aleiner}}\ and\ \bibinfo {author} {\bibfnamefont {K.~A.}\ \bibnamefont
  {Matveev}},\ }\href {\doibase 10.1103/PhysRevLett.80.814} {\bibfield
  {journal} {\bibinfo  {journal} {Phys. Rev. Lett.}\ }\textbf {\bibinfo
  {volume} {80}},\ \bibinfo {pages} {814} (\bibinfo {year} {1998})}\BibitemShut
  {NoStop}%
\bibitem [{\citenamefont {Gefen}\ \emph {et~al.}(2002)\citenamefont {Gefen},
  \citenamefont {Berkovits}, \citenamefont {Lerner},\ and\ \citenamefont
  {Altshuler}}]{Gefen-2002}%
  \BibitemOpen
  \bibfield  {author} {\bibinfo {author} {\bibfnamefont {Y.}~\bibnamefont
  {Gefen}}, \bibinfo {author} {\bibfnamefont {R.}~\bibnamefont {Berkovits}},
  \bibinfo {author} {\bibfnamefont {I.~V.}\ \bibnamefont {Lerner}}, \ and\
  \bibinfo {author} {\bibfnamefont {B.~L.}\ \bibnamefont {Altshuler}},\ }\href
  {\doibase 10.1103/PhysRevB.65.081106} {\bibfield  {journal} {\bibinfo
  {journal} {Phys. Rev. B}\ }\textbf {\bibinfo {volume} {65}},\ \bibinfo
  {pages} {081106} (\bibinfo {year} {2002})}\BibitemShut {NoStop}%
\bibitem [{\citenamefont {Nandkishore}\ and\ \citenamefont
  {Huse}(2015)}]{Nandkishore-2015}%
  \BibitemOpen
  \bibfield  {author} {\bibinfo {author} {\bibfnamefont {R.}~\bibnamefont
  {Nandkishore}}\ and\ \bibinfo {author} {\bibfnamefont {D.~A.}\ \bibnamefont
  {Huse}},\ }\href@noop {} {\bibfield  {journal} {\bibinfo  {journal} {Annu.
  Rev. Condens. Matter Phys.}\ }\textbf {\bibinfo {volume} {6}},\ \bibinfo
  {pages} {15} (\bibinfo {year} {2015})}\BibitemShut {NoStop}%
\bibitem [{\citenamefont {Khemani}\ \emph {et~al.}(2015)\citenamefont
  {Khemani}, \citenamefont {Nandkishore},\ and\ \citenamefont
  {Sondhi}}]{Khemani-2015}%
  \BibitemOpen
  \bibfield  {author} {\bibinfo {author} {\bibfnamefont {V.}~\bibnamefont
  {Khemani}}, \bibinfo {author} {\bibfnamefont {R.}~\bibnamefont
  {Nandkishore}}, \ and\ \bibinfo {author} {\bibfnamefont {S.~L.}\ \bibnamefont
  {Sondhi}},\ }\href@noop {} {\bibfield  {journal} {\bibinfo  {journal} {Nat.
  Phys.}\ }\textbf {\bibinfo {volume} {11}},\ \bibinfo {pages} {560} (\bibinfo
  {year} {2015})}\BibitemShut {NoStop}%
\bibitem [{\citenamefont {Deng}\ \emph {et~al.}(2015)\citenamefont {Deng},
  \citenamefont {Pixley}, \citenamefont {Li},\ and\ \citenamefont
  {Das~Sarma}}]{Deng-2015}%
  \BibitemOpen
  \bibfield  {author} {\bibinfo {author} {\bibfnamefont {D.-L.}\ \bibnamefont
  {Deng}}, \bibinfo {author} {\bibfnamefont {J.~H.}\ \bibnamefont {Pixley}},
  \bibinfo {author} {\bibfnamefont {X.}~\bibnamefont {Li}}, \ and\ \bibinfo
  {author} {\bibfnamefont {S.}~\bibnamefont {Das~Sarma}},\ }\href {\doibase
  10.1103/PhysRevB.92.220201} {\bibfield  {journal} {\bibinfo  {journal} {Phys.
  Rev. B}\ }\textbf {\bibinfo {volume} {92}},\ \bibinfo {pages} {220201}
  (\bibinfo {year} {2015})}\BibitemShut {NoStop}%
\bibitem [{\citenamefont {Deng}\ \emph {et~al.}(2017)\citenamefont {Deng},
  \citenamefont {Li}, \citenamefont {Pixley}, \citenamefont {Wu},\ and\
  \citenamefont {Das~Sarma}}]{Deng-2017}%
  \BibitemOpen
  \bibfield  {author} {\bibinfo {author} {\bibfnamefont {D.-L.}\ \bibnamefont
  {Deng}}, \bibinfo {author} {\bibfnamefont {X.}~\bibnamefont {Li}}, \bibinfo
  {author} {\bibfnamefont {J.~H.}\ \bibnamefont {Pixley}}, \bibinfo {author}
  {\bibfnamefont {Y.-L.}\ \bibnamefont {Wu}}, \ and\ \bibinfo {author}
  {\bibfnamefont {S.}~\bibnamefont {Das~Sarma}},\ }\href {\doibase
  10.1103/PhysRevB.95.024202} {\bibfield  {journal} {\bibinfo  {journal} {Phys.
  Rev. B}\ }\textbf {\bibinfo {volume} {95}},\ \bibinfo {pages} {024202}
  (\bibinfo {year} {2017})}\BibitemShut {NoStop}%
\bibitem [{\citenamefont {Vasseur}\ and\ \citenamefont
  {Moore}(2015)}]{vasseur_moore}%
  \BibitemOpen
  \bibfield  {author} {\bibinfo {author} {\bibfnamefont {R.}~\bibnamefont
  {Vasseur}}\ and\ \bibinfo {author} {\bibfnamefont {J.~E.}\ \bibnamefont
  {Moore}},\ }\href {\doibase 10.1103/PhysRevB.92.054203} {\bibfield  {journal}
  {\bibinfo  {journal} {Phys. Rev. B}\ }\textbf {\bibinfo {volume} {92}},\
  \bibinfo {pages} {054203} (\bibinfo {year} {2015})}\BibitemShut {NoStop}%
\bibitem [{\citenamefont {Knap}\ \emph {et~al.}(2012)\citenamefont {Knap},
  \citenamefont {Shashi}, \citenamefont {Nishida}, \citenamefont {Imambekov},
  \citenamefont {Abanin},\ and\ \citenamefont {Demler}}]{Knap-2012}%
  \BibitemOpen
  \bibfield  {author} {\bibinfo {author} {\bibfnamefont {M.}~\bibnamefont
  {Knap}}, \bibinfo {author} {\bibfnamefont {A.}~\bibnamefont {Shashi}},
  \bibinfo {author} {\bibfnamefont {Y.}~\bibnamefont {Nishida}}, \bibinfo
  {author} {\bibfnamefont {A.}~\bibnamefont {Imambekov}}, \bibinfo {author}
  {\bibfnamefont {D.~A.}\ \bibnamefont {Abanin}}, \ and\ \bibinfo {author}
  {\bibfnamefont {E.}~\bibnamefont {Demler}},\ }\href {\doibase
  10.1103/PhysRevX.2.041020} {\bibfield  {journal} {\bibinfo  {journal} {Phys.
  Rev. X}\ }\textbf {\bibinfo {volume} {2}},\ \bibinfo {pages} {041020}
  (\bibinfo {year} {2012})}\BibitemShut {NoStop}%
\bibitem [{\citenamefont {Azbel}(1979)}]{Azbel-1979}%
  \BibitemOpen
  \bibfield  {author} {\bibinfo {author} {\bibfnamefont {M.~Y.}\ \bibnamefont
  {Azbel}},\ }\href {\doibase 10.1103/PhysRevLett.43.1954} {\bibfield
  {journal} {\bibinfo  {journal} {Phys. Rev. Lett.}\ }\textbf {\bibinfo
  {volume} {43}},\ \bibinfo {pages} {1954} (\bibinfo {year}
  {1979})}\BibitemShut {NoStop}%
\bibitem [{\citenamefont {Aubry}\ and\ \citenamefont
  {Andr{\'e}}(1980)}]{Aubry-1980}%
  \BibitemOpen
  \bibfield  {author} {\bibinfo {author} {\bibfnamefont {S.}~\bibnamefont
  {Aubry}}\ and\ \bibinfo {author} {\bibfnamefont {G.}~\bibnamefont
  {Andr{\'e}}},\ }\href@noop {} {\bibfield  {journal} {\bibinfo  {journal}
  {Ann. Israel Phys. Soc}\ }\textbf {\bibinfo {volume} {3}},\ \bibinfo {pages}
  {133} (\bibinfo {year} {1980})}\BibitemShut {NoStop}%
\bibitem [{\citenamefont {Roati}\ \emph {et~al.}(2008)\citenamefont {Roati},
  \citenamefont {D’Errico}, \citenamefont {Fallani}, \citenamefont {Fattori},
  \citenamefont {Fort}, \citenamefont {Zaccanti}, \citenamefont {Modugno},
  \citenamefont {Modugno},\ and\ \citenamefont {Inguscio}}]{Roati-2008}%
  \BibitemOpen
  \bibfield  {author} {\bibinfo {author} {\bibfnamefont {G.}~\bibnamefont
  {Roati}}, \bibinfo {author} {\bibfnamefont {C.}~\bibnamefont {D’Errico}},
  \bibinfo {author} {\bibfnamefont {L.}~\bibnamefont {Fallani}}, \bibinfo
  {author} {\bibfnamefont {M.}~\bibnamefont {Fattori}}, \bibinfo {author}
  {\bibfnamefont {C.}~\bibnamefont {Fort}}, \bibinfo {author} {\bibfnamefont
  {M.}~\bibnamefont {Zaccanti}}, \bibinfo {author} {\bibfnamefont
  {G.}~\bibnamefont {Modugno}}, \bibinfo {author} {\bibfnamefont
  {M.}~\bibnamefont {Modugno}}, \ and\ \bibinfo {author} {\bibfnamefont
  {M.}~\bibnamefont {Inguscio}},\ }\href@noop {} {\bibfield  {journal}
  {\bibinfo  {journal} {Nature}\ }\textbf {\bibinfo {volume} {453}},\ \bibinfo
  {pages} {895} (\bibinfo {year} {2008})}\BibitemShut {NoStop}%
\bibitem [{\citenamefont {Combescot}\ and\ \citenamefont
  {Nozi{\`e}res}(1971)}]{combescot1971infrared}%
  \BibitemOpen
  \bibfield  {author} {\bibinfo {author} {\bibfnamefont {M.}~\bibnamefont
  {Combescot}}\ and\ \bibinfo {author} {\bibfnamefont {P.}~\bibnamefont
  {Nozi{\`e}res}},\ }\href@noop {} {\bibfield  {journal} {\bibinfo  {journal}
  {Journal de Physique}\ }\textbf {\bibinfo {volume} {32}},\ \bibinfo {pages}
  {913} (\bibinfo {year} {1971})}\BibitemShut {NoStop}%
\bibitem [{\citenamefont {Hofstadter}(1976)}]{Hofstadter-1976}%
  \BibitemOpen
  \bibfield  {author} {\bibinfo {author} {\bibfnamefont {D.~R.}\ \bibnamefont
  {Hofstadter}},\ }\href {\doibase 10.1103/PhysRevB.14.2239} {\bibfield
  {journal} {\bibinfo  {journal} {Phys. Rev. B}\ }\textbf {\bibinfo {volume}
  {14}},\ \bibinfo {pages} {2239} (\bibinfo {year} {1976})}\BibitemShut
  {NoStop}%
\bibitem [{\citenamefont {Hiramoto}\ and\ \citenamefont
  {Kohmoto}(1992)}]{hiramoto1992electronic}%
  \BibitemOpen
  \bibfield  {author} {\bibinfo {author} {\bibfnamefont {H.}~\bibnamefont
  {Hiramoto}}\ and\ \bibinfo {author} {\bibfnamefont {M.}~\bibnamefont
  {Kohmoto}},\ }\href@noop {} {\bibfield  {journal} {\bibinfo  {journal} {Int.
  J. Mod. Phys. B}\ }\textbf {\bibinfo {volume} {6}},\ \bibinfo {pages} {281}
  (\bibinfo {year} {1992})}\BibitemShut {NoStop}%
\bibitem [{\citenamefont {Mahan}(2013)}]{Mahan-2013}%
  \BibitemOpen
  \bibfield  {author} {\bibinfo {author} {\bibfnamefont {G.~D.}\ \bibnamefont
  {Mahan}},\ }\href@noop {} {\emph {\bibinfo {title} {Many-particle physics}}}\
  (\bibinfo  {publisher} {Springer Science \& Business Media, New York},\
  \bibinfo {year} {2013})\BibitemShut {NoStop}%
\bibitem [{\citenamefont {Szab\'o}\ and\ \citenamefont
  {Schneider}(2018)}]{Schneider-2018}%
  \BibitemOpen
  \bibfield  {author} {\bibinfo {author} {\bibfnamefont {A.}~\bibnamefont
  {Szab\'o}}\ and\ \bibinfo {author} {\bibfnamefont {U.}~\bibnamefont
  {Schneider}},\ }\href {\doibase 10.1103/PhysRevB.98.134201} {\bibfield
  {journal} {\bibinfo  {journal} {Phys. Rev. B}\ }\textbf {\bibinfo {volume}
  {98}},\ \bibinfo {pages} {134201} (\bibinfo {year} {2018})}\BibitemShut
  {NoStop}%
\bibitem [{\citenamefont {Cao}\ \emph {et~al.}(2018)\citenamefont {Cao},
  \citenamefont {Fatemi}, \citenamefont {Fang}, \citenamefont {Watanabe},
  \citenamefont {Taniguchi}, \citenamefont {Kaxiras},\ and\ \citenamefont
  {Jarillo-Herrero}}]{Cao-2018}%
  \BibitemOpen
  \bibfield  {author} {\bibinfo {author} {\bibfnamefont {Y.}~\bibnamefont
  {Cao}}, \bibinfo {author} {\bibfnamefont {V.}~\bibnamefont {Fatemi}},
  \bibinfo {author} {\bibfnamefont {S.}~\bibnamefont {Fang}}, \bibinfo {author}
  {\bibfnamefont {K.}~\bibnamefont {Watanabe}}, \bibinfo {author}
  {\bibfnamefont {T.}~\bibnamefont {Taniguchi}}, \bibinfo {author}
  {\bibfnamefont {E.}~\bibnamefont {Kaxiras}}, \ and\ \bibinfo {author}
  {\bibfnamefont {P.}~\bibnamefont {Jarillo-Herrero}},\ }\href@noop {}
  {\bibfield  {journal} {\bibinfo  {journal} {Nature}\ }\textbf {\bibinfo
  {volume} {556}},\ \bibinfo {pages} {43} (\bibinfo {year} {2018})}\BibitemShut
  {NoStop}%
\bibitem [{\citenamefont {Gaunt}\ \emph {et~al.}(2013)\citenamefont {Gaunt},
  \citenamefont {Schmidutz}, \citenamefont {Gotlibovych}, \citenamefont
  {Smith},\ and\ \citenamefont {Hadzibabic}}]{zoran2013}%
  \BibitemOpen
  \bibfield  {author} {\bibinfo {author} {\bibfnamefont {A.~L.}\ \bibnamefont
  {Gaunt}}, \bibinfo {author} {\bibfnamefont {T.~F.}\ \bibnamefont
  {Schmidutz}}, \bibinfo {author} {\bibfnamefont {I.}~\bibnamefont
  {Gotlibovych}}, \bibinfo {author} {\bibfnamefont {R.~P.}\ \bibnamefont
  {Smith}}, \ and\ \bibinfo {author} {\bibfnamefont {Z.}~\bibnamefont
  {Hadzibabic}},\ }\href {\doibase 10.1103/PhysRevLett.110.200406} {\bibfield
  {journal} {\bibinfo  {journal} {Phys. Rev. Lett.}\ }\textbf {\bibinfo
  {volume} {110}},\ \bibinfo {pages} {200406} (\bibinfo {year}
  {2013})}\BibitemShut {NoStop}%
\end{thebibliography}%

\end{document}